\documentclass[aps,superscriptaddress,nofootinbib,showpacs]{revtex4}
\usepackage{hyperref}
\usepackage{epsfig,rotating}
\usepackage{amsmath,amssymb}
\usepackage{dsfont}

\numberwithin{equation}{section}

\newcommand{\vx}{\vec{x}}
\newcommand{\vp}{\vec{p}}

\newcommand{\vk}{\vec{k}}

\newcommand{\al}{\alpha}

\newcommand{\ga}{\gamma}

\newcommand{\uvk}{\widehat{\bf{k}}}

\newcommand{\be}{\begin{equation}}
\newcommand{\ee}{\end{equation}}
\newcommand{\bea}{\begin{eqnarray}}
\newcommand{\eea}{\end{eqnarray}}

\begin{document}

\title{Sterile neutrinos produced near the EW scale I:\\ mixing angles, MSW resonances and production rates.}
\author{Jun Wu}\email{juw31@pitt.edu}  \affiliation{Department of Physics and
Astronomy, University of Pittsburgh, Pittsburgh, PA 15260}
\author{Chiu Man Ho} \email{cmho@berkeley.edu}
\affiliation{Department of Physics, University of California,
Berkeley, CA 94720, USA} \affiliation{Theoretical Physics Group,
Lawrence Berkeley National Laboratory, Berkeley, CA 94720}
\author{Daniel Boyanovsky}
\email{boyan@pitt.edu} \affiliation{Department of Physics and
Astronomy, University of Pittsburgh, Pittsburgh, PA 15260}

\date{\today}

\begin{abstract}
We study the production of sterile neutrinos in the region $T \sim
M_W$   in an extension beyond the standard model with the see-saw
mass matrix  originating  in Yukawa couplings to Higgs-like scalars
with masses and vev's of the order of the electroweak scale. Sterile
neutrinos are produced by the decay of
  scalars and  standard model vector bosons. We obtain the index of refraction, dispersion relations,
  mixing angles in the medium and production rates including those for right-handed sterile neutrinos,
  from the standard model and beyond the standard model
  self-energies. For $1 \lesssim M_W/T \lesssim 3$ we find narrow MSW resonances with $k \lesssim T$ for both
  left and right handed neutrinos even in absence of a
  lepton asymmetry in the (active) neutrino   sector,  as well as very low energy  ($k/T \ll |\xi|$)
   narrow MSW resonances in the presence of a lepton asymmetry consistent   with the bounds from WMAP and BBN.  For
     small vacuum mixing angle, consistent with observational bounds,  the absorptive part of the self-energies
     lead  to a strong damping regime   very near the resonances resulting in  the \emph{exact} degeneracy
     of the propagating modes with a concomitant breakdown of   adiabaticity. We argue that cosmological
     expansion
  sweeps through the resonances,   \emph{resonant and non-resonant} sterile neutrino production results in  a
  highly \emph{non-thermal} distribution function enhanced at small momentum $k < T$,
    with potentially important consequences for their free streaming length  and transfer function at
    small scales.

\end{abstract}


\maketitle

\section{Introduction}\label{sec:intro}
In the concordance $\Lambda\mathrm{CDM}$ standard cosmological
model, the Universe today is composed approximately by $70\%$ of a
dark energy component responsible for the acceleration, about $25\%$
of dark matter (DM) and about $5\%$ of ordinary matter (baryons). In
this scenario the (DM) component is cold and collisionless, and
structure formation proceeds in a hierarchical ``bottom-up'' manner:
small scales become non-linear and collapse first and their merger
and accretion lead to structures on larger scales\cite{primack}.
This is a consequence of the fact that cold dark matter (CDM)
particles feature negligible velocity dispersion leading to a power
spectrum that favors small scales. In this hierarchical scenario
dense clumps that survive the merger process form satellite
galaxies. Numerical simulations of structure formation with (CDM)
predict many orders of magnitude more (DM) sub-haloes   than
observed low luminosity dwarf
galaxies\cite{klypin,moore2,kauff,moore,diemand,vialactea}. These
simulations also yield a density profile that increases
monotonically towards the
center\cite{frenk,moore2,dubi,bullock,cusps,vialactea} $\rho(r) \sim
r^{-\gamma}$, $\gamma=1$ corresponds to the Navarro-Frenk-White
(NFW) profile, but steeper profiles with $\gamma \sim 1.2$ have been
found recently in numerical simulations\cite{vialactea}. These
density profiles accurately describe clusters of galaxies but there
has been recent observational evidence that seems to indicate a
shallow cored profile instead of cusps in dwarf spheroidal galaxies
(dShps) which are deemed to be (DM)
dominated\cite{kleyna,dalcanton1,gilmore,salucci,mcgaugh,battaglia}.
This core vs. cusp controversy is still being debated, and recent
arguments suggest that the interpretation of the data is subject to
assumptions and modelling\cite{evans}. Recently yet another
discrepancy between the predictions of $\Lambda$CDM and observations
has been revealed, the ``emptiness of voids'', possibly related to
the overabundance problem\cite{klypin0807}.

Warm dark matter particles (WDM) were
invoked\cite{mooreWDM,turokWDM,avila,knebe} as possible solutions to
the core vs. cusps and the overabundance problems in satellite galaxies.
(WDM) particles feature a non-vanishing velocity dispersion with a
range in between (CDM) and hot dark matter (HDM)  leading to a free
streaming scale that cutoffs power at small scales thereby smoothing
out  small scale structure. If the free streaming scale of the (WDM)
particles is smaller than the scale of galaxy clusters, the large
scale structure properties are indistinguishable from those of
(CDM), but may affect structure at small scales\cite{bond}, thereby
providing an explanation of the smoother inner profiles and the
fewer satellites. A small scale cutoff in the (DM) power spectrum
may also explain the apparent smallness of galaxies at $z\sim 3$
found in ref.\cite{cen}.

 Although the interpretation of cores in
(dSphs) may be challenged by alternative explanations, and the
missing satellite problem could be resolved by astrophysical
mechanisms such as complex ``gastrophysics'', and   recent
simulations suggest that the dynamics of subhalos is not too
different in (WDM) and (CDM) models\cite{knebe2}, there is an
intrinsic interest in studying alternatives to the standard (CDM)
paradigm.

Any particle physics explanation of (DM) involves extensions beyond
the Standard Model (SM),     allowing  quite generally, both  (CDM)
and (WDM) candidates.

Sterile neutrinos, namely $SU(2)$ singlets,  with masses in the
$\sim \mathrm{keV}$ range may be suitable (WDM)
candidates\cite{dw,colombi,este,shapo,kusenko,kusepetra,petra,coldmatter,micha}
and may provide possible solutions to other astrophysical
problems\cite{kusenko}. The main property that is relevant for
structure formation of any (DM) candidate is its distribution
function after decoupling\cite{hogan,coldmatter} which depends on
the production mechanism and the quantum kinetic evolution from
production to decoupling. Sterile neutrinos may be produced by
various different
mechanisms\cite{dw,colombi,este,shapo,kusenko,kusepetra,petra},
among them non-resonant mixing, or Dodelson-Widrow
(DW)\cite{dw,colombi,este} has been invoked often. However, there
seems to be some tension between the X-ray\cite{xray} and the
Lyman-$\alpha$ forest data\cite{lyman,lyman2} leading to the
suggestion\cite{palazzo}  that (DW)-produced sterile neutrinos
cannot be the dominant (WDM) component.

A phenomenologically appealing extension of the (SM) with only one
scale has been recently proposed\cite{shapo,asaka,micha}. In this
model  sterile neutrinos may be produced by the decay of a
gauge-singlet scalar with a mass of the order of the electroweak
scale\cite{kusenko,kusepetra,petra,kuse2}. In this scenario sterile
neutrinos  are produced and decouple at a temperature of the  order
of the   mass or the scalar\cite{kusepetra,petra,boysnudm}.

Recently\cite{boysnudm} the quantum kinetics of production and
decoupling of $\sim \mathrm{keV}$ sterile neutrinos in these models
was studied with the result that production via the decay of the
gauge-singlet scalar leads to a non-thermal distribution function
that favors small momentum. This result was combined with an
analytic method to obtain the transfer function during matter
domination recently introduced in ref.\cite{freestream1}. This
method
 reveals the influence of the distribution function of the
decoupled particle upon the power spectrum and free streaming
length\cite{freestream2}. The results of ref.\cite{boysnudm} point
out that sterile neutrinos produced via the decay of gauge-singlet
scalars in the model advocated in
refs.\cite{shapo,kusenko,kusepetra,petra,kuse2,boysnudm} yield
\emph{smaller} free streaming lengths and an  \emph{enhancement} of
power at small scales as compared to those produced by the (DW)
mechanism. Combining the results for the distribution function of
sterile neutrinos produced via scalar decay with abundance and phase
space constraints from dwarf spheroidal galaxies
(dShps)\cite{coldmatter} yields a narrow window for the mass of
sterile neutrinos\cite{boysnudm}: $ 0.56 ~ \mathrm{keV} \lesssim M_s
\lesssim 1.33 ~\mathrm{keV}$. The robustness of this bound has been
confirmed in ref.\cite{hector}, but there may be some tension with recent analysis of the Lyman-$\alpha$ forest with non-thermal populations\cite{newboyarski}, although the
results in this reference relate mainly to resonant production.

Recent observations of the X-ray spectra from the Ursa Minor dwarf
spheroidal galaxy with the \emph{Suzako} satellite\cite{kuseobs}
suggest that sterile neutrinos with masses in the  $\mathrm{keV}$
range with mixing angles $\theta \sim 10^{-5}$ remain viable
candidates as main dark matter constituents, a result that seems to
be confirmed by  those of ref.\cite{newboyarski}.

\vspace{2mm}

\textbf{Motivation and objectives:} The clustering properties of
dark matter candidates depend on the free streaming length which
determines the scale below which power is suppressed. When the (DM)
particle of mass $M_s$ has become non-relativistic, the free
streaming length is approximately given by
\[ \lambda_{fs} \simeq \Big[ \langle p^2 \rangle/M^2_s G \rho \Big]^\frac{1}{2}, \]
 where  $\rho $ is the (DM) density and the average is with the distribution
function of the decoupled (DM) particle. Distribution functions that favor small momenta lead to smaller
free streaming lengths and more power a small scales\cite{boysnudm,freestream1,freestream2}.

 The study in ref.\cite{boysnudm} revealed that
the \emph{non-resonant} mechanism of sterile neutrino production by
scalar decay advocated in ref.\cite{kusepetra,petra,kuse2,shapo}
leads to a non-thermal distribution function that favors small
momenta with important consequences for structure formation and
remarkable differences with sterile neutrinos produced by the (DW)
mechanism\cite{dw}, whose distribution function is that of a thermal
relic that decoupled while relativistic, but
multiplied by an overall factor\cite{dw}. This overall factor in the
(DW) distribution function \emph{only} affects the abundance, but
for a fixed (DM) density $\rho$   the resulting free streaming
length is that of a neutrino of mass $M_s$ decoupled at the sterile
neutrino decoupling temperature. For a fixed mass and relic density
the non-thermal distribution function from the production mechanism
studied in ref.\cite{boysnudm,kusepetra,petra} yields a
\emph{smaller} $\lambda_{fs}$ and more power at
 small scales than in the (DW) mechanism without modifying the large scale power spectrum.

In the extension beyond the standard model (bsm)  advocated in
ref.\cite{kusepetra,petra,kuse2,shapo,boysnudm} sterile neutrinos
\emph{mix} with active neutrinos via a Yukawa coupling to the
standard model Higgs\cite{shapo} whose expectation value yields a
see-saw mass matrix. The diagonalization of this see-saw mass matrix
yields  interaction vertices between vector bosons  and the sterile-like neutrino. This is important because the
distribution functions being a function of the energy, are
necessarily associated with  mass or energy eigenstates, \emph{not}
flavor eigenstates.

The study in ref.\cite{kusepetra,petra,boysnudm} reveals that
sterile neutrinos are produced and decouple at a temperature of the
order of the mass of the scalar, which in the model of
refs.\cite{shapo,kusepetra,petra,kuse2} is of the order of the Higgs
mass.

At this temperature the charged and neutral vector bosons are
present in the medium with large abundance, comparable to that of
the scalar. \emph{Their decay into the sterile-like neutrinos will
therefore contribute to their total abundance and distribution
function.} \emph{This is   one of the main observations in this
article.} The coupling of the charged and neutral vector bosons to
the sterile-like neutrino is suppressed by the (small) mixing angle,
but since the standard model couplings are much larger than the
Yukawa couplings of the scalar to the sterile neutrino, the question
is whether the decay of vector bosons may lead to a substantial
contribution to the production rate of the sterile-like neutrinos.
For $M_s \sim \mathrm{keV}$ and the expectation value of the
Higgs-like scalar in the range $\sim 100\,\mathrm{GeV}$ the Yukawa
coupling $Y \sim 10^{-8}$, the production rate via this process
$\propto Y^2$, whereas the contribution from $Z,W$ decay would be
expected to be $\propto \alpha_w \sin^2(\theta)$, with  $\theta$ the
mixing angle. For $\theta \sim 10^{-5}$\cite{kuse2,kuseobs} the
production rate of sterile-like neutrinos via  vector boson decay
can be of the same order of \emph{or larger than} that from scalar
decay.   This observation suggests that sterile neutrino production
via the decay of \emph{vector bosons} in the medium may be
competitive with the production via scalar decay.

At high temperature and or density the mixing angle is modified by
medium effects\cite{msw,book1,book2,book3,notzold,boyhonueu,dolgov},
therefore the first step towards understanding whether vector boson
decay contributes to the   production of sterile-like neutrinos is to
obtain the in-medium correction to the mixing angles.

A more general   aspect of   sterile neutrino
production via vector boson decay at $T\sim M_W$  is that both the
index of refraction (real part of the self-energy) and the
production rate determined by the absorptive part (imaginary part of
the self-energy) are of $\mathcal{O}(G_F)$. This is in contrast to
the usual situation at temperatures much smaller than the
electroweak scale when the index of refraction is of
$\mathcal{O}(G_F)$, but the absorptive part is of
$\mathcal{O}(G^2_F)$.

Although the finite temperature and density corrections to the index
of refraction have been obtained for $T \ll
M_{W,Z}$\cite{notzold,boyhonueu,dolgov}, to the best of our knowledge
the study of the self-energy, the index of refraction (real part)
and absorptive part (width) at $T \simeq M_{W,Z}$ has not been
carried out.

To be sure, upon the diagonalization of the mass matrix, standard
model interaction vertices with the sterile-like neutrino lead to
production processes via both charged and neutral current
interactions such as $\overline{l}l \rightarrow \overline{\nu}_1
\nu_2~; \overline{f} f \rightarrow \overline{\nu}_1 \nu_2$ with charged leptons (l) or quarks (f) and $\nu_1 \sim \nu_a;\nu_2\sim \nu_s$ (active and sterile respectively) for
small mixing angle. These processes are
  of $\mathcal{O}(\alpha^2_w \sin^2 \theta )$ and while they will eventually become important for $T \ll
M_{W,Z}$ when the population of vector bosons in the medium becomes
$\ll \alpha_w$, these
  are formally subleading in
the weak coupling at $T \sim M_{W,Z}$.

Therefore at $T\sim M_{W,Z}$ vector boson decay is the
\emph{leading} production mechanism from weak interactions.

Our   objective is to provide a comprehensive assessment of sterile
neutrinos as potential (DM) candidates implementing the following
program:

\begin{itemize}
\item{Obtain the production rates and  mixing angle in the medium from the
 quantum field theory model at $T \sim M_W$,
studying  the possibility of MSW resonances to determine whether
 sterile neutrinos are produced resonantly or non-resonantly.   }

 \item{Obtain  and solve the kinetic equations describing the production and
 decoupling of sterile neutrinos using the rates and mixing angles
 obtained from previous step.}

 \item{ The asymptotic solution of the kinetic equation yields the
 distribution function after freeze out, which determines the abundance and the free streaming length. This
 distribution function is input in the program described in refs.\cite{freestream1,boysnudm} to obtain the transfer
 function and power spectrum.}

\end{itemize}

In this article we carry out the first step of this program. We
implement methods of field theory at finite temperature and density
developed in
refs.\cite{boyaneutrino,boyhonueu,hoboya,hoboydavey,nosfermions} to
obtain the mixing angles in the medium and production rates both
from scalar and vector boson decay.

\medskip

\textbf{Results:} We study a simple extension of the standard model
with one active and one sterile neutrino to extract the robust
features in a simpler setting. Both active and sterile neutrinos are
considered to be Dirac, this is to include the possibility of a
lepton asymmetry hidden in the (active) neutrino sector (Majorana
neutrinos cannot be assigned a chemical potential), and to allow us to study the
production of left and right-handed neutrinos.

We obtain the dispersion relations, index of refraction, mixing
angles and production rates in the medium from the self-energy
contributions from standard model (sm) and beyond the standard model (bsm)
interactions.
 The see-saw mass matrix that mixes them emerges from the Yukawa couplings to Higgs-like scalars
with masses of the order of $M_{W,Z}$ that acquire expectation values also of this order. We focus on the temperature region $T \sim M_{W}$ where   vector and scalar bosons are present in the medium with
large thermal populations. The decay of both the scalar and vector bosons contribute to the production
of sterile neutrinos. Our main results are:

\begin{itemize}
\item{We find one MSW resonance even in \emph{absence} of a lepton asymmetry. For $ 1 \lesssim M_W/T
 \lesssim 3$ this resonance is in the low momentum region $ 0.2 \lesssim k/T \lesssim 1$ and well within the regime
 of validity of the perturbative expansion. Including a lepton asymmetry
    in the active neutrino sector consistent with the data from Wilkinson Microwave Anisotropy Probe (WMAP)
    and Big Bang Nucleosynthesis (BBN)\cite{steigman}, we find
    \emph{two} low energy MSW resonances, the lowest one is a consequence of the lepton asymmetry that occurs
    at $k/T \ll \xi$ with $|\xi|$ being the lepton asymmetry parameter. In the region of interest for this
    study for small vacuum mixing angle consistent with the observational
    bounds from X-ray data\cite{kuseobs}
    these resonances are very narrow. We find resonances also for positive energy, positive helicity, namely nearly \emph{right-handed}
    neutrinos.  }

    \item{ At the resonances the propagating frequencies become exactly \emph{degenerate} in striking
     contrast with the quantum mechanics of neutrino mixing wherein there is level repulsion at the
     resonance. This exact degeneracy at the resonance entails the breakdown of adiabaticity. It is a
     distinct consequence of the absorptive part of the self-energy and leads to a strong damping regime. }

\item{The form of the standard model contribution to the production rate is similar to that from
scalar decay found in ref.\cite{boysnudm}. We argue that
cosmological expansion will lead to a rapid crossing of the narrow
resonances resulting in both \emph{resonant and non-resonant}
sterile neutrino production. In particular nearly \emph{right
handed} sterile neutrinos are produced by the decay of $Z^0,W^{\pm}$
vector bosons. Their distribution functions after freeze-out will be
\emph{highly non-thermal} with a distinct enhancement at small
momentum $k< T$ and perturbatively small population.   This low
momentum enhancement of the non-thermal distribution function is
expected\cite{freestream1,boysnudm} to have important consequences:
a shortening of the free-streaming length (smaller velocity
dispersion) and an increase of the transfer function and power
spectrum at small scales.   }

\item{We find a consistent range of parameters for which there is a resonance for positive helicity, positive
energy neutrinos, namely nearly right-handed at $T \sim M_W$. The
general field theory framework allows a systematic study of the
properties for both helicity states, including the helicity
dependence of mixing angles and production rates.    }

\end{itemize}

\section{The Model} \label{sec:model}
The extension of the standard model presented in
ref.\cite{kusepetra,petra,kuse2}   generalizes the proposal of the
$\nu$-MSM of ref.\cite{asaka,shapo} and is also a generalization of
the model presented in ref.\cite{degouvea}. These models include
three SU(2) singlet (sterile) neutrinos which couple to the active
neutrinos via a see-saw mass matrix. The generalization of
ref.\cite{kusepetra,petra,kuse2} gives a mass to the sterile
neutrino via a Yukawa coupling to a Higgs-like scalar field which
\emph{could} be the neutral Higgs component, or another scalar whose
expectation value is of the same order as that of the (sm) Higgs boson, therefore this type
of extension features only one scale.

We study a simplified version of these models by considering only
one sterile and one active neutrino. In the usual see-saw mechanism an off-diagonal
Dirac mass matrix for the active species is considered along with a diagonal Majorana
 mass for the sterile neutrino\cite{asaka,shapo,book1,book2,book3}. However, instead of
considering a Majorana sterile neutrino, we allow for  Dirac mass
terms for all species. This generalization allows to study simultaneously the possibility of
a lepton asymmetry in the (active) neutrino sector for which a Dirac field is required,
 along with the possibility of a right-handed component leading  to
potentially relevant degrees of freedom within the same simple
model.

Our goal is to extract   generic and robust features of the
production rates and mixing angles in the medium along with a
reliable estimate of sterile production rates. The generalization to
three species can be done relatively straightforwardly (but for the
complications associated with dealing with larger mixing matrices),
and the case of a Majorana neutrino is regained straightforwardly by
projection.

We consider a model with one active ($\nu_a$) and one sterile ($\nu_s$) (an $SU(2)$ singlet) Dirac neutrinos, described by the Lagrangian density

\bea\label{Lag} \mathcal{L} = \mathcal{L}_{SM}+\overline{\nu}_s \,i
{\not\!{\partial}}\,  \, \nu_s
 -Y_1 ~ \overline{\nu}_s \tilde{H}^\dagger l -Y_2 ~\overline{\nu}_s \Phi \nu_s + \mathcal{L}[\Phi]+ \mathrm{h.c} ~,\eea where \be l = \Bigg( \begin{array}{c}
                                          \nu_a \\
                                          f
                                        \end{array}
  \Bigg) ~~;~~ \tilde{H} = \Bigg( \begin{array}{c}
                                          H^0 \\
                                          H^-
                                        \end{array}
  \Bigg) \,.\label{doublets}\ee $f$ is the charged lepton associated with $\nu_a$ and $H^0,H^-$ are
  the  components of the standard model Higgs doublet, and $\Phi$ is a real scalar
singlet field whose expectation value gives a Dirac mass to the
sterile neutrino. The Lagrangian density
 $\mathcal{L}[\Phi]$ describes the kinetic and potential terms of
 $\Phi$.

  In unitary gauge we write\be H^0 = \langle H^0 \rangle +\sigma ~~;~~ \Phi = \langle \Phi \rangle + \varphi \label{vevs}\ee and consistently with the single scale assumption of the $\nu$-MSM: $\langle H^0 \rangle \sim \langle \Phi \rangle$
are of the same order of magnitude (the weak scale ) and that their masses are also of the same scale. In fact our analysis is quite general, and  this assumption will only be
invoked for a quantitative assessment. The Lagrangian density (\ref{Lag}) becomes

 \bea\label{Laguni} \mathcal{L} = \mathcal{L}_{SM}+\overline{\nu}_s   \,i
{\not\!{\partial}} \nu_s - \overline{\nu}_\alpha ~\mathds{M}_{\alpha \beta}~  \nu_{\beta}
 -Y_1 ~ \overline{\nu}_s \sigma \nu_a -Y_2 ~ \overline{\nu}_s \varphi \nu_s +
 \mathcal{L}[\langle \Phi \rangle + \varphi]+ \mathrm{h.c}~; ~\alpha,\beta = a,s ~,\eea where

\be \label{massmatrix} \mathds{M}=\left(
\begin{array}{cc}
 0 & m   \\
  m   & M_s \\
\end{array}%
\right)~~;~~ m=Y_1~\langle H^0 \rangle ~~;~~ M_s = Y_2~ \langle \Phi
\rangle. \ee Introducing the ``flavor'' doublet $( {\nu}_a,
 {\nu}_s)$ the diagonalization
 of the mass term $\mathds{M}$ is achieved by a unitary transformation to the mass basis
  $( {\nu}_1,  {\nu}_2)$, namely \be \Big( \begin{array}{c}
                                                          \nu_a \\
                                                          \nu_s
                                                        \end{array}
\Big) = U(\theta) \Big( \begin{array}{c}
                                                          \nu_1 \\
                                                          \nu_2
                                                        \end{array} \Big) ~~;~~
                                                         U(\theta) = \Bigg( \begin{array}{cc}
                                                                              \cos(\theta) & \sin(\theta) \\
                                                                              -\sin(\theta) & \cos(\theta)
                                                                            \end{array}
                                                          \Bigg) ~,\label{unitrafo}\ee
where \be \cos(2\theta) = \frac{M_s}{\left[M^2_s + 4 m^2 \right]^\frac{1}{2}} ~~;~~   \sin(2\theta) = \frac{2 m}{\left[M^2_s + 4 m^2 \right]^\frac{1}{2}}~.
 \label{mixingangles}\ee In the mass basis \be  {\mathds{M}}_m= U^{-1}(\theta)~;~\mathds{M} U(\theta) = \Bigg( \begin{array}{cc}
                                                M_1 & 0 \\
                                               0 & M_2
                                              \end{array}
 \Bigg)~~;~~M_1 = \frac{1}{2}\left[M_s - \left[M^2_s + 4 m^2 \right]^\frac{1}{2}\right] ~~;~~
  M_2 = \frac{1}{2}\left[M_s + \left[M^2_s + 4 m^2 \right]^\frac{1}{2}\right]~. \label{diagmass}\ee

We focus on a see-saw with $M_s \sim \mathrm{keV} \gg m$ therefore
\be M_1 \simeq -\frac{m^2}{M_s} ~~;~~M_2 \simeq M_s ~~;~~
\sin(2\theta) \simeq \frac{2m}{M_s} \sim
\left|\frac{M_1}{M_2}\right|^\frac{1}{2} \ll 1 ~.\label{parameters}\ee

Taking $\langle H^0 \rangle \sim \langle \Phi \rangle$ the small
mixing angle entails that $Y_1 \ll Y_2$ which   results in
self-energy corrections from the $\sigma$ exchange are   subleading
as compared to those from the $\varphi$ exchange. For example taking
$\langle \Phi \rangle \sim \langle H^0 \rangle$,   and for a $\sim
\mathrm{keV}$ sterile neutrino it follows that \be Y_2 \sim 10^{-8}
\gg Y_1 ~~;~~ \sin(2\theta) \sim  {Y_1}/{Y_2}\,.\label{Ys}\ee

However, we can   alternatively   consider  a
pre-determined see-saw mass matrix and set $Y_1=Y_2=0$  which
corresponding to a simpler extension of the standard model that posits
a mass matrix that originates beyond the standard model.

Our goal is to obtain the dynamical aspects of sterile neutrinos in
the medium, mixing angles, dispersion relations and damping rates,
which determine the production rates. These are obtained directly
from the solution of the equations of motion including the
self-energy corrections in the medium. The one-loop self-energies
require the neutrino \emph{propagators} in the medium in the
\emph{mass basis}, since the mass eigenstates are the true
propagating states. For $\theta \ll 1$ the mass eigenstates $\nu_1
\sim \nu_a; \nu_2 \sim \nu_s$, and the active neutrino reaches
equilibrium at $T \gtrsim 1\, \mathrm{MeV}$ via the weak
interactions, whereas the sterile neutrinos are not expected to
equilibrate.

This argument, however, hinges on the smallness of the vacuum mixing
angle, but in a medium the mixing angle can become very large, and if
there are MSW resonances the roles of the medium eigenstates may be
reversed. Whether there are MSW resonances and the medium mixing angle becomes
large can only be answered \emph{a posteriori}.

Therefore we \emph{assume} that    the mass
eigenstate $\nu_1$ is active-like, and features a Fermi-Dirac distribution function,
whereas for $\nu_2$ the propagators are the vacuum ones.
Furthermore, it is possible that if there is a large lepton
asymmetry it may be stored in the neutrino sector, whereas the
asymmetry in the charged leptons equals the baryon asymmetry and can
be neglected. Hence the Fermi-Dirac distribution functions in the
$\nu_1$ propagator includes a chemical potential.

In our study we   explicitly separate the fermionic and bosonic contributions to the
self-energies to assess the consistency of   the assumption that the
eigenstate ``1'' is active-like.

\section{Equations of motion}\label{sec:real-time}

The   effective Dirac equation in the medium is derived  with the methods of non-equilibrium quantum field theory
described in \cite{hoboya,boyaneutrino,nosfermions}. We follow  the
approach presented in refs.\cite{hoboya,boyaneutrino}   and introduce an
external Grassmann-valued source that couples linearly to the
neutrino field via the Lagrangian density

\be \mathcal{L}_S = \overline{\nu}_\al  \; \eta_\al +
\overline{\eta}_\al \; \nu_\al \; , \label{Lsource} \ee

\noindent whence the total lagrangian density is given by
$\mathcal{L}+\mathcal{L}_S$. The external source induces an
expectation value for the neutrino field which   obeys the
effective equation of motion with self-energy corrections from the
  medium \cite{nosfermions}.

The equation of motion   is derived by shifting
the field $ \nu^{\pm}_\al = \psi_\al + \Psi^{\pm}_\al \, , \, \psi_\al
=  \langle \nu^{\pm}_\al \rangle  $    imposing $\langle \Psi^{\pm}_\al \rangle =0$ order by
order in the perturbation theory
\cite{nosfermions,boyaneutrino,hoboya}. Since the self-energy
  corrections to the equations of motion require the
neutrino propagators, we obtain the equation of motion in the mass
basis.

Implementing this program up to one loop order, we find the
following equation of motion for the doublet in the mass basis $
\psi^T  \equiv \big( \psi_1 \,,\,  \psi_2\big)$, it is given by
\be \left(\,i\not\!{\partial}\,\mathds{I}-
{\mathds{M}_m}+\Sigma^{tad}_{sm}\,L\,\right)\,\psi (\vx,t) + \int
d^3 x'   dt' \; \left[\,\Sigma^{ret}_{sm}(\vx-\vx',t-t')
\;L+\Sigma^{ret}_{bsm}(\vx-\vx',t-t')\,\right] \psi(\vx',t') = -
\eta(\vx,t), \label{eqnofmot} \ee

\noindent where $\mathds{I}$ is the identity matrix, $\mathds{M}_m= diag(M_1,M_2)$
is the mass matrix in the mass basis,  $L=(1-\gamma^5)/2$
is the left-handed chiral projection operator, $\Sigma^{tad}_{sm}$
is the (local) tadpole contribution from the (sm) neutral
current interaction, (see Fig. (\ref{fig:smloop})) $.
\Sigma^{ret}_{sm}(\vx-\vx',t-t')$ and
$\Sigma^{ret}_{bsm}(\vx-\vx',t-t')$ are respectively the real-time
retarded self-energies   from    (sm) and   (bsm) (scalar) interactions. Introducing the
space-time Fourier transform in a spatial volume $V$ \be
\psi(\vec{x},t) = \frac{1}{\sqrt{V}} \sum_{\vec{k}}\int dk_0
e^{i\vec{k}\cdot\vec{x}}\,e^{-ik_0t} \tilde{\psi}(k_0, \vec{k})
\label{FT}\ee and similarly for the self-energy kernels and the
source term, the equation of motion in the mass basis becomes \be
\Bigg[\big(\gamma_0 k_0 - \vec{\gamma}\cdot\vec{k}\big)\mathds{I} -
 {\mathds{M}_m} + \Sigma^{tad}_{sm}\,L + \Sigma_{sm}(k_0,\vec{k}) \,L +
\Sigma_{bsm}(k_0,\vec{k}\,) \Bigg]\tilde{\psi}(k_0,\vk) = - \tilde{\eta}(k_0,\vk)~.
\label{eqnofmotFT}\ee

The space-time Fourier transform of the retarded self-energies (not
the tadpole) feature a dispersive representation

\be \Sigma(k_0,k) = \frac{1}{\pi}\int_{-\infty}^\infty d\omega
\; \frac{\mathrm{Im}\Sigma(\omega,\vec{k}\,)}{\omega-k_0 -i\,0^+ } \;
 .\label{selfa}\ee

\subsection{One-Loop Self-Energy}\label{sec:self-energy}

We   focus on the temperature region  $M_{Z,W,\sigma,\varphi}
\gtrsim T$, in which using the unperturbed thermal propagators for
the scalar and vector bosons is valid\cite{htl}.  In section ({\ref{subsec:PT}) we show that
 perturbation theory is valid for $k \gtrsim \alpha_w\,T$ for
 $M_{W,\sigma,\varphi} \gtrsim T $, furthermore for $k \ll M_{W} $   our results reproduce those found in the literature for $T\ll M_{W}$\cite{notzold,boyhonueu} and the perturbative     expansion is reliable for $M_{W} \gtrsim 2 T$.

The (sm) charged and neutral current contributions   to the
self-energy in the mass basis are depicted in
Fig.(\ref{fig:smloop}). The latin indices $i,j,k=1,2$ refer to the
mass basis fields and the label $f$ in the intermediate fermion
propagator in the charged current diagram in Fig.(\ref{fig:smloop})
refers to the charged lepton associated with the active neutrino.
The contributions from scalar exchange (bsm) in the mass basis are
depicted in Fig.(\ref{fig:bsmloop}).

\begin{figure}
\begin{center}
\includegraphics[height=6cm,width=8cm,keepaspectratio=true]{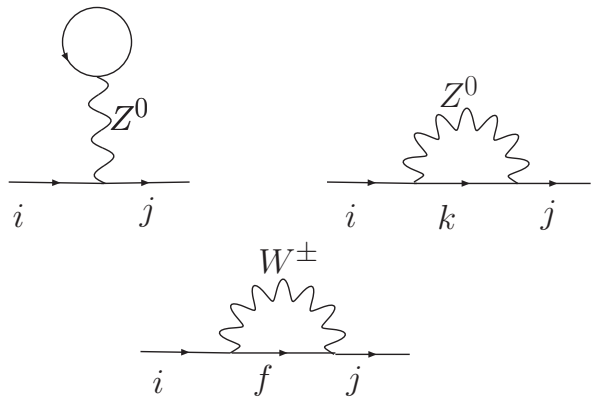}
\caption{Standard model contributions   to the  self-energy $\Sigma_{sm}$. The indices $i,k,j=1,2$ corresponding to mass eigenstates,
the index $f$ for the intermediate fermion line in the charged-current self-energy refers to the
charged lepton associated with the active neutrino.}
\label{fig:smloop}
\end{center}
\end{figure}

\begin{figure}
\begin{center}
\includegraphics[height=5cm,width=6cm,keepaspectratio=true]{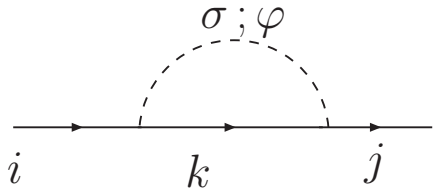}
\caption{ Beyond the standard model contributions   to the  self-energy $\Sigma_{bsm}$.
The indices $i,k,j=1,2$ corresponding to mass eigenstates. The dashed line is a scalar propagator either for $\sigma$
or $\varphi$}
\label{fig:bsmloop}
\end{center}
\end{figure}

{\bf (SM) neutral currents:} The tadpole contribution in the mass
basis is given by \be \Sigma^{tad}_{sm} = \Sigma^{t}~~
U^{-1}(\theta)\left(
                \begin{array}{cc}
                  1 & 0 \\
                  0 & 0 \\
                \end{array}
              \right)
    U(\theta)~,  \label{sigmatad} \ee where\footnote{This expression corrects a typographic error in ref.\cite{hoboya}).} \be \Sigma^t = - \gamma^0
\frac{g^2}{4 M^2_W}\int
\frac{d^3q}{(2\pi)^3}\big(n_\nu-\overline{n}_\nu\big) = - \gamma^0
\frac{g^2 T^3}{24 M^2_W}\,\xi \Big[1+\frac{\xi^2}{\pi^2}\Big] ~~;~~
\xi = \frac{\mu}{T}\,. \label{sigta} \ee  In this expression $n_\nu,
\overline{n}_\nu$ are the Fermi-Dirac distribution functions for
neutrinos and antineutrinos respectively, and  we have neglected the
contribution from the asymmetry of the charged lepton and quark
sectors since these are proportional to the (negligible) baryon
asymmetry. We allow for a lepton asymmetry stored in the neutrino
sector.  A recent analysis\cite{steigman} from the latest WMAP and
BBN data suggests that $|\xi| \lesssim 10^{-2}$.

The neutral current diagrams that contribute to the one-loop self
energy feature two different terms corresponding to the intermediate
neutrino line being either $\nu_1$ or $\nu_2$. As argued above, for
small mixing angles $\nu_1 \sim \nu_a$ and weak interactions
equilibrate these mass eigenstates with the medium, therefore their
finite temperature propagator features the Fermi-Dirac distribution
function (with a chemical potential allowing for a lepton
asymmetry). However, $\nu_2 \sim \nu_s$ will \emph{not} equilibrate
with the medium since their coupling to the environmental degrees of
freedom is suppressed by at least two powers of the (small) mixing
angle, therefore   $\nu_2$ features  a vacuum propagator. The one
loop diagrams are shown in Fig. (\ref{fig:ncse}) where  the superscripts $(1)$ and $(2)$ are used to
specify the intermediate neutrino propagator $\nu_1$ and $\nu_2$
respectively.

\begin{figure}
\begin{center}
\includegraphics[height=5cm,width=8cm,keepaspectratio=true]{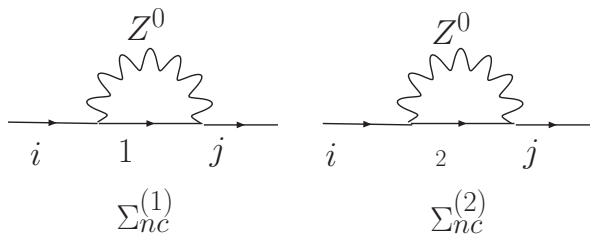}
\caption{ Neutral currents contribution to the one-loop retarded
self-energy $\Sigma_{sm}$. The indices $i,j =1,2$ and the  indices $
1,2$ denote the corresponding mass eigenstate in the intermediate
state. } \label{fig:ncse}
\end{center}
\end{figure}

In the mass basis we find for the neutral current contributions
shown in Fig. (\ref{fig:ncse}) \be \Sigma_{nc}(k_0,\vk) =
\Big[\cos^2(\theta)\,\Sigma^{(1)}_{nc}(k_0,\vk) +
\sin^2(\theta)\,\Sigma^{(2)}_{nc}(k_0,\vk) \Big]
\,U^{-1}(\theta)\left(
                \begin{array}{cc}
                  1 & 0 \\
                  0 & 0 \\
                \end{array}
              \right)
    U(\theta)~. \label{sigmanc} \ee

{\bf (sm) charged currents:} the charged current one-loop self
energy is shown in Fig. (\ref{fig:smloop}), since the intermediate
state is a charged lepton we find in the mass basis  \be
\Sigma_{cc}(k_0,\vk) = \Sigma_{cc,sm}(k_0,\vk)
\,U^{-1}(\theta)\left(
                \begin{array}{cc}
                  1 & 0 \\
                  0 & 0 \\
                \end{array}
              \right)
    U(\theta) \;,\label{sigmanc} \ee where $ \Sigma_{cc,sm}(k_0,\vk)$
    is the usual standard model one-loop self-energy in thermal equilibrium.

   {\bf (bsm) scalar exchange:} The scalar exchange contributions to
   the self-energy are shown in Fig.(\ref{fig:bsmloop2}). For
   $\sin^2(\theta) \ll 1$ we find \bea \Sigma_{bsm}(k_0,\vk) = &&
   \Bigg[\cos^2(\theta)\,\Sigma^{(1)}_{\sigma}(k_0,\vk) +
   \sin^2(\theta)\Sigma^{(1)}_{\varphi}(k_0,\vk)+\cos^2(\theta)\Sigma^{(2)}_{\varphi}(k_0,\vk)
   \Bigg]\,U^{-1}(\theta)\left(
                \begin{array}{cc}
                  0 & 0 \\
                  0 & 1 \\
                \end{array}
              \right)
    U(\theta) + \nonumber \\ && \cos^2(\theta)\,\Sigma^{(2)}_{\sigma}(k_0,\vk)\,U^{-1}(\theta)\left(
                \begin{array}{cc}
                  1 & 0 \\
                  0 & 0 \\
                \end{array}
              \right)  U(\theta)~. \label{sigmabsmcont}\eea

\begin{figure}
\begin{center}
\includegraphics[height=5cm,width=10cm,keepaspectratio=true]{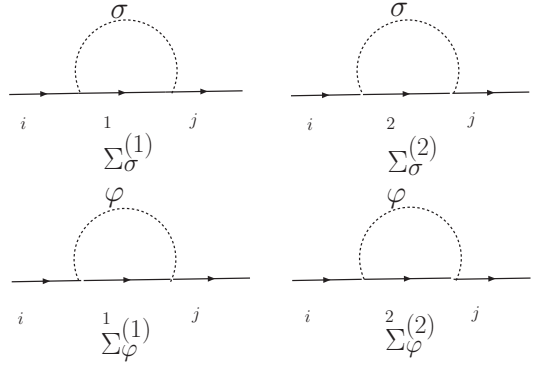}
\caption{ Scalar exchange contributions to the one-loop self-energy
$\Sigma_{bsm}$. The indices $i,j =1,2$ and the indices $ 1,2$ denote
the corresponding mass eigenstate in the intermediate state. }
\label{fig:bsmloop2}
\end{center}
\end{figure}

\textbf{Summary of self-energies in the flavor basis:} The structure
of the self-energies (to leading order in $\sin^2(\theta)$)
(\ref{sigmatad}-\ref{sigmabsmcont}) indicates that they are
 {diagonal} in the \emph{flavor} basis. In this basis the total self-energy
is given by \be \Sigma(k_0,\vk) = \Bigg( \begin{array}{cc}
                                                \Sigma_{aa}(k_0,\vk) & 0 \\
                                                0 &
                                                \Sigma_{ss}(k_0,\vk)
                                              \end{array}
  \Bigg)~,
\label{totsigfla}\ee where \bea \Sigma_{aa}(k_0,\vk) & = &
\Big[\Sigma^t + \cos^2(\theta)\,\Sigma^{(1)}_{nc}(k_0,\vk) +
\sin^2(\theta)\,\Sigma^{(2)}_{nc}(k_0,\vk)+ \Sigma_{cc,sm}(k_0,\vk)
\Big]L
+ \cos^2(\theta)\,\Sigma^{(2)}_{\sigma}(k_0,\vk)~, \label{sigmaaa} \\
\Sigma_{ss}(k_0,k) & = &
\cos^2(\theta)\,\Sigma^{(1)}_{\sigma}(k_0,\vk) +
   \sin^2(\theta)\Sigma^{(1)}_{\varphi}(k_0,\vk)+\cos^2(\theta)\Sigma^{(2)}_{\varphi}(k_0,\vk)~. \label{sigmass}\eea

Since in the (sm) contributions we have explicitly factored out the
left-handed projector  $L$, the remainder contributions to the (sm)
self-energies $\Sigma_{nc,cc}$ are those of a vector-like theory. The
(bsm) contributions feature both chiralities since we have
considered a Dirac mass term for the sterile neutrino, a left-handed
Majorana mass term can be obtained by neglecting the right-handed
contribution.  We   consider  the regime   $T\gg M_{1,2},m_f$  and  $k_0, k \gg M_{1,2},m_f$, where
$m_f$ stand for the charged lepton masses, therefore we can safely
neglect the mass terms and consider the propagators of massless
fermionic fields.

In this regime   the general form
of the (sm)  self-energies with vector boson exchange, either
charged or neutral currents is written in dispersive form as in eqn.
(\ref{selfa}) with\cite{hoboya,boyaneutrino}

\be Im\Sigma_{sm}(\omega,\vk) =   \pi  { g^2_{sm} } ~\int
\frac{d^3q}{(2\pi)^3} \int dp_0\,dq_0 \delta(\omega-p_0-q_0)
\Big[1-n_{F}(p_0)+N_B(q_0)\Big] \gamma^\mu
\rho_F(p_0,\vec{p})\rho_B(q_0,\vec{q})\gamma^\nu\,P_{\mu
\nu}(q_0,\vec{q})~, \label{imsigsm}\ee where $F$ stands for the
fermionic species in the intermediate state. For $\nu_1$ and charged lepton $n_F$ is the
Fermi-Dirac distribution function, whereas for $\nu_2$ it is $n_F=0$
since the ``sterile'' neutrino does not thermalize with the medium.
For the (bsm) contributions, the general form for scalar exchange is
\be Im\Sigma_{bsm}(\omega,\vk) = \pi Y^2\int \frac{d^3q}{(2\pi)^3}
\int dp_0\,dq_0 \delta(\omega-p_0-q_0)
\Big[1-n_{F}(p_0)+N_B(q_0)\Big]
\rho_F(p_0,\vec{p})\rho_B(q_0,\vec{q})~. \label{imsigbsm}\ee where
  \be g_{sm} = \Bigg\{ \begin{array}{c}
                                          \displaystyle{\frac{g}{\sqrt{2}}} ~~~~~~~~\mathrm{CC} \\
                                         \displaystyle{\frac{g}{ {2}\cos(\theta_w)}}
                                         ~~\mathrm{NC}
                                        \end{array} \label{gsm}\ee and  $Y=Y_1,Y_2$ for $\sigma$ and $\varphi$
                                     exchange respectively. The
spectral densities are respectively (for massless fermions) \bea
\rho_F(p_0,\vec{p})  & = & \frac{1}{2}\Big( \gamma^0 - \vec{\gamma}\cdot
\frac{\vec{p}}{p}\Big)~\delta(p_0-p)+ \frac{1}{2}\Big( \gamma^0 +
\vec{\gamma}\cdot \frac{\vec{p}}{p}\Big)~\delta(p_0+p)~,
\label{ferspecdens}\\
\rho_B(q_0,\vec{q}) & = &
\frac{1}{2W_q}\big[\delta(q_0-W_q)-\delta(q_0+W_q)\big] ~~;~~W_q =
\sqrt{q^2+M^2}~. \label{bosspecdens}\eea The projection operator \be
P_{\mu \nu}(q_0,\vec{q}) = -\Bigg[g_{\mu \nu}- \frac{q_\mu
q_\nu}{M^2_{Z,W}} \Bigg]~~;~~q^\mu = (q^0, \vec{q})
\label{projector} \ee and \be n_{F}(p_0)    =
\frac{1}{e^{(p_0-\mu)/T}+1} ~~;~~ \overline{n}_F(p_0) =
1-n_\nu(-p_0)~~;   N_B(q_0)    =  \frac{1}{e^{q_0/T}-1}~.
\label{DFs} \ee  We have allowed a chemical potential for the
neutrinos (only for $\nu_1 \sim \nu_a$) to include the possibility
of a   lepton asymmetry in the (active) neutrino sector.

In the expressions above, the masses for the scalars or vector
bosons are $M_{\sigma,\varphi},M_{Z,W}$ as appropriate for each
contribution. All the self-energies   share the general form \be \Sigma(k_0,\vec{k}) \equiv  \gamma^0 \, {A}(k_0,k
)-\vec{\ga}\,\cdot\uvk ~ {B}(k_0, k)\,, \label{generalsig}\ee the detailed expressions for the
  imaginary parts of the (sm) and (bsm) contributions are given in the appendices.

 In particular, for the neutral current tadpole $B(k_0,k)=0$ and $A(k_0,k)$ can be recognized from
 eqn. (\ref{sigta}).  Combining (\ref{totsigfla}) with this form
we write the self-energy matrix in the \emph{flavor basis} as

\be  \Sigma^{tad}_{sm}\,L + \Sigma_{sm}(k_0,\vec{k}) \,L +
\Sigma_{bsm}(k_0,\vec{k}\,) \equiv \Bigg[\gamma^0 ~\mathds{A}_L(k_0,k )-
\vec{\gamma}\cdot\uvk ~\mathds{B}_L(k_0,k )\Bigg]\,L +\Bigg[\gamma^0 ~
\mathds{A}_R(k_0,k )- \vec{\gamma}\cdot\uvk ~\mathds{B}_R(k_0,k
)\Bigg]\,R \,.\label{sigmatrix} \ee

In the flavor basis these
matrices   are of the form   \be
\mathds{A}(k_0,k) = \Bigg(\begin{array}{cc}
                               A_{aa}(k_0,k) & 0 \\
                               0 & A_{ss}(k_0,k)
                             \end{array} \Bigg)~~;~~\mathds{B}(k_0,k)   =   \Bigg(\begin{array}{cc}
                               B_{aa}(k_0,k) & 0 \\
                               0 & B_{ss}(k_0,k)
                             \end{array} \Bigg)\,, \label{ABmatx}
                             \ee where the matrix elements are obtained  from the
expressions (\ref{sigmaaa},\ref{sigmass}).

The equations of motion for the  left (L) and right (R) handed
components are obtained by multiplying the equation of motion (\ref{eqnofmotFT}) on the left by the
projectors $R$ and $L$ respectively.

It proves convenient at this
stage to separate the Dirac spinors into the left $\psi_L$ and right
$\psi_R$ handed components and to expand them into helicity
eigenstates\cite{hoboya}, namely  \be \psi_L = \sum_{h=\pm 1} v^{h}
\otimes \varphi^h ~~;~~\varphi^h = \Bigg( \begin{array}{c}
                                            \varphi^h_a \\
                                            \varphi^h_s
                                          \end{array}
\Bigg)~, \label{varfiL} \ee and \be \psi_R = \sum_{h=\pm 1} v^{h}
\otimes  \zeta^h ~~;~~\zeta^h = \Bigg( \begin{array}{c}
                                            \zeta^h_a \\
                                            \zeta^h_s
                                          \end{array}
\Bigg)~, \label{varfiR} \ee where the left $\varphi $ and right
$\zeta$  handed doublets are written in the \emph{flavor} basis, and
$v^h$ are eigenstates of the  helicity operator \be
\widehat{h}(\uvk) = \gamma^0 \vec{\gamma}\cdot\,\uvk\,\gamma^5 =
\vec{\sigma}\cdot\,\uvk ~\Bigg( \begin{array}{cc}
                                  \mathds{1} & 0 \\
                                  0 & \mathds{1}
                                \end{array}
\Bigg)\label{helicity}\ee namely, \be \vec{\sigma}\cdot\,\uvk ~ v^h
= h ~ v^h~~;~~h=\pm 1 \,.\label{vhel}\ee

To leading order in weak and Yukawa couplings, and  neglecting a
commutator $[\mathds{M},\Sigma]$ because it is higher order in these
couplings, we find in the \emph{flavor} basis for both  the left and
right-handed component doublets  \be \Bigg[ ( k^2_0 - k^2)
\mathds{I} + \big(k_0-h k\big)\big(\mathds{A}_L + h
\mathds{B}_L\big)+\big(k_0+h k\big)\big(\mathds{A}_R - h
\mathds{B}_R\big) -\mathds{M}^2 \Bigg]\Bigg\{ \begin{array}{c}
                \varphi^h \\
                \zeta^h
              \end{array}
 \Bigg\}  = \Bigg\{ \begin{array}{c}
                I^h_L \\
                I^h_R
              \end{array}
 \Bigg\}   \,, \label{leftrighteom}\ee where $\mathds{M}$ is the mass matrix in the
\emph{flavor} basis and the inhomogeneities in these equations are
obtained by projection and using the corresponding equations, we
need not specify them as they are no longer used in our study.

In absence of interactions, for the left-handed component  a
positive energy solution corresponds to $h=-1$ and a negative energy
solution to $h=+1$ with the opposite assignment for the right-handed
component.

In the flavor basis \be  {\mathds{M}}^2 = {\overline{M}}^{\,2}
\mathds{I} + \frac{\delta M^2}{2}  \, \Bigg(
                            \begin{array}{cc}
                             -\cos(2\theta) & \sin(2\theta) \\
                              \sin(2\theta) & \cos(2\theta) \\
                            \end{array}
                            \Bigg) \,.\label{M2flav}\ee where \be \overline{M}^{\,2} \equiv
                            \frac{1}{2}\big(M^2_1+M^2_2\big)~~;~~\delta
                            M^2 \equiv  M^2_2-M^2_1  \,,
                            \label{massdefs}\ee and $M_{1,2}$ are
  given by eqn. (\ref{diagmass}).

                            It proves convenient to define the combinations
\bea S_h(k_0,k) & = & (k_0+h k)\Big[ \big(\mathds{A}_R
-h\mathds{B}_R \big)_{aa} + \big(\mathds{A}_R-h\mathds{B}_R
\big)_{ss}  \Big]    \nonumber \\ & + & (k_0-h k)\Big[
\big(\mathds{A}_L+h\mathds{B}_L \big)_{aa} +
\big(\mathds{A}_L+h\mathds{B}_L \big)_{ss}  \Big]~, \label{S} \eea and
\bea \Delta_h(k_0,k) & = & \frac{(k_0+h k)}{\delta M^2 }\Big[
\big(\mathds{A}_R-h\mathds{B}_R \big)_{aa} -
\big(\mathds{A}_R-h\mathds{B}_R \big)_{ss}  \Big]    \nonumber \\ &
+ & \frac{(k_0-h k)}{\delta M^2}\Big[
\big(\mathds{A}_L+h\mathds{B}_L \big)_{aa} -
\big(\mathds{A}_L+h\mathds{B}_L \big)_{ss}  \Big]~, \label{Delta} \eea
where we have suppressed the arguments.
The equation of motion (\ref{leftrighteom}) can now be written as \be \mathds{G}^{-1}_h(k_0,k)\Bigg\{ \begin{array}{c}
                \varphi^h \\
                \zeta^h
              \end{array}
 \Bigg\}  = \Bigg\{ \begin{array}{c}
                I^h_L \\
                I^h_R
              \end{array}
 \Bigg\}~, \label{Seom}\ee where the inverse propagator is given by \be \mathds{G}^{-1}_h(k_0,k) =
 \Big(k^2_0-k^2 +\frac{1}{2}S_h(k_0,k)- \overline{M}^{\,2}\Big)\mathds{I}-\frac{1}{2}\delta M^2 \rho_h(k_0,k) \,
 \Bigg(
   \begin{array}{cc}
     -C_h(k_0,k)  & D_h(k_0,k) \\
   D_h(k_0,k) & C_h(k_0,k) \\
   \end{array}
 \Bigg)~, \label{invS}\ee  where
 \be \rho_h(k_0,k)= \Bigg[\Big(\cos(2\theta)+  {\Delta_h(k_0,k)}  \Big)^2 +\sin^2(2\theta)\Bigg]^\frac{1}{2} \label{rho} \ee and
 \bea C_h(k_0,k)&  =  & \frac{\Big(\cos(2\theta)+ {\Delta_h(k_0,k)} \Big) }{\rho_h(k_0,k)}~, \label{cos2tetam} \\
 D_h(k_0,k)&  =  & \frac{\sin(2\theta) }{\rho_h(k_0,k)}~. \label{sin2tetam} \eea

 We note that \emph{if} $\Delta_h(k_0,k)$   were real, then $C_h(k_0,k) = \cos(2\theta^h_m(k_0,k))$ and $D_h(k_0,k) = \sin(2\theta^h_m(k_0,k))$
 with
  $\theta^h_m(k_0,k)$   the mixing angle in the medium for the different
 helicity projections and as a function of frequency and momentum.

 \subsection{Propagator: complex poles and propagating modes in the
 medium}

 From (\ref{invS}) we read off the propagator projected onto helicity eigenstates
 \be \mathds{G}_h(k_0,k) = \frac{\mathds{I}+\mathds{T}_h(k_0,k)}{2\Big(\alpha_h(k_0,k)-\beta_h(k_0,k)\Big)} +
 \frac{\mathds{I}-\mathds{T}_h(k_0,k)}{2\Big(\alpha_h(k_0,k)+\beta_h(k_0,k)\Big)}~,  \label{propah} \ee where
 \be \mathds{T}_h(k_0,k) = \Bigg(
   \begin{array}{cc}
     -C_h(k_0,k) & D_h(k_0,k)  \\
   D_h(k_0,k)  & C_h(k_0,k)  \\
   \end{array}
 \Bigg)~, \label{CDmat}\ee

 \bea \alpha_h(k_0,k) & =  & k^2_0-k^2 +\frac{1}{2}S_h(k_0,k)- \overline{M}^{\,2}~, \label{alfa}\\
 \beta_h(k_0,k) & = & \frac{1}{2}\delta M^2 \rho_h(k_0,k)~. \label{beta} \eea If $\Delta_h(k_0,k)$ given by eqn. (\ref{Delta}) \emph{were} real,
  the propagator (\ref{propah})
  would be diagonalized by the unitary transformation \be U_h(\theta^h_m(k_0,k)) =
 \left(
   \begin{array}{cc}
      \cos( \theta^h_m(k_0,k)) & \sin(\theta^h_m(k_0,k)) \\
     -\sin(\theta^h_m(k_0,k)) & \cos(\theta^h_m(k_0,k)) \\
   \end{array}
 \right)\,, \label{Um} \ee leading to   \be U^{-1} (\theta_m)\,\mathds{G}(k_0,k)
 U(\theta_m) = \left(
                   \begin{array}{cc}
                     \displaystyle{\frac{1}{\alpha(k_0,k)+\beta(k_0,k)}} & 0 \\
                    0 &  \displaystyle{\frac{1}{\alpha(k_0,k)-\beta(k_0,k)} }  \\
                   \end{array}
                 \right)~, \label{Sdiag}\ee where we have suppressed the helicity argument
for simplicity. However, because $\Delta_h(k_0,k)$ features an
imaginary part determined by the absorptive part  of the
self-energies, there is no unitary transformation that diagonalizes
the propagator. However, since the imaginary part is perturbatively
small the expression (\ref{Sdiag}) clearly indicates that the pole
for $\alpha = \beta$ corresponds to the mass eigenstate $2$, namely
a sterile-like neutrino state, and the pole for $\alpha = -\beta$
corresponds to the mass eigenstate $1$, namely an active-like state.

We   note that in absence of interactions, namely $S_h
=0~;~\Delta_h =0$ it follows that \bea \alpha+\beta & = &
k^2_0-k^2-M^2_1~, \label{alplusbe0} \\ \alpha-\beta & = &
k^2_0-k^2-M^2_2 ~. \label{alminbe0}\eea

The propagating eigenstates in the medium are determined by the (complex) poles of the propagator (\ref{propah}),
 which again correspond to $\alpha_h(k_0,k) = \pm \beta_h(k_0,k)$.

Before we analyze the complex poles, it proves convenient to
separate the real and imaginary parts of $\alpha,\beta$. For this
purpose and to simplify notation, we suppress the label $h$ and the
arguments $k_0,k$ in these quantities, and we write \be S = S_R + i
S_I ~~;~~\Delta= \Delta_R + i\Delta_I~, \label{reim}\ee where the
subscripts $R,I$ stand for real and imaginary parts respectively.
Furthermore, we \emph{define} the mixing angles in the medium solely
in terms of the \emph{real} parts of the self-energy (index of
refraction), namely \be  \cos(2\theta_m)    =
\frac{\cos(2\theta)+\Delta_R}{\rho_0}~~;~~\sin(2\theta_m) =
\frac{\sin(2\theta)}{\rho_0}  \,, \label{mediumangle}\ee where \be
\rho_0 = \Bigg[\Big(\cos(2\theta)+  {\Delta_R}  \Big)^2
+\sin^2(2\theta)\Bigg]^\frac{1}{2}\,. \label{rho0}\ee

An MSW resonance occurs whenever $\cos(2\theta_m)=0$\cite{msw,book1,book2,book3}, namely when
 \be \Delta_R = -\cos(2\theta) \,. \label{MSWcond}\ee

We emphasize
that the mixing angle in the medium $\theta_m$ and $\rho_0$ depend
on \emph{helicity}, $k_0,k$. In terms of these quantities we find
\be \beta = \frac{\delta M^2}{2}\rho_0\,r\,\big[\cos(\phi) + i
\sin(\phi)\big]  \equiv \beta_R + i\beta_I~, \label{betRI}\ee
where
 \be r = \Bigg[\big(1-\tilde{\gamma}^2 \big)^2+ \big(2\tilde{\gamma}\cos(2\theta_m) \big)^2 \Bigg]^\frac{1}{4}
 ~~;~~ \tilde{\gamma} = \frac{\Delta_I}{\rho_0}~, \label{ar}\ee
 and \be \phi = \mathrm{sign}\big(\tilde{\gamma} \cos(2\theta_m) \big)\Bigg\{ \frac{1}{2} \mathrm{arctg}\Bigg|\frac{2\tilde{\gamma} \cos(2\theta_m) }
 {1-\tilde{\gamma}^2}  \Bigg|~\Theta(1-\tilde{\gamma}^2) + \Bigg(\frac{\pi}{2}-  \frac{1}{2} \mathrm{arctg}\Bigg|\frac{2\tilde{\gamma}
 \cos(2\theta_m) }{1-\tilde{\gamma}^2}  \Bigg| \Bigg) ~\Theta(\tilde{\gamma}^2-1)
 \Bigg\}\,.
 \label{alfaangle}\ee This form is similar to that obtained in a
  model of oscillations and damping with mixed neutrinos studied in
 ref.\cite{boymodel}, and suggests two distinct situations: a \emph{weak damping} case for
  $|\tilde{\gamma}| < 1$ and a \emph{strong damping} case for $|\tilde{\gamma}| >1$. These cases
  will be analyzed below.

 \textbf{Zeroes of $\alpha + \beta$:} We are concerned with the ultrarelativistic limit $k \gg M^2_2 \gg M^2_1$. Just as in the usual case\cite{book1,book2,book3} it is convenient to introduce the
  average or reference frequency \be \overline{\omega}(k) =
  \sqrt{k^2+\overline{M}^{\,2}}\,.
   \label{overomega}\ee The poles are near $\overline{\omega}(k)$, therefore
    write   \be k_0 =   \overline{\omega}(k) + \big(k_0
 -  \overline{\omega}  (k)\big)   \,, \label{k0} \ee   keeping only
 the linear term in $ \big(k_0
 - {\overline{\omega}}(k)\big) $ we find \be  \alpha+\beta \sim
 2 \overline{\omega}(k)\Bigg[k_0 -  {\Omega}_1(k) + i \Gamma_1(k)  \Bigg]
 \label{alplusbe}\ee
with \bea {\Omega}_1(k) &  = & \overline{\omega}(k)
 -\frac{1}{4 \overline{\omega}(k)} \Bigg[ S_R + \delta M^2  \rho_0 r
 \cos(\phi) \Bigg]_{k_0 =  \overline{\omega}(k)}~, \label{shiftR1}\\ \Gamma_1(k) &
 = & \frac{1}{4 \overline{\omega}(k)} \Bigg[ S_I + \delta M^2 \rho_0 r
 \sin(\phi)\Bigg]_{k_0 =  \overline{\omega}(k)}~. \label{shiftI1}
 \eea

 \textbf{Zeroes of $\alpha-\beta$:} Proceeding in the same manner,  we find

 \be  \alpha-\beta \sim
 2 \overline{\omega}(k)\Bigg[k_0 - \Omega_2(k) + i \Gamma_2(k)  \Bigg]
 \label{alminbe}\ee with \bea \Omega_2(k) &  = & \overline{\omega}(k)
 -\frac{1}{4 \overline{\omega}(k)} \Bigg[ S_R - \delta M^2 \rho_0 r
 \cos(\phi) \Bigg]_{k_0 =  \overline{\omega}(k)}~, \label{shiftR2}\\ \Gamma_2(k) &
 = & \frac{1}{4 \overline{\omega}(k)} \Bigg[ S_I - \delta M^2 \rho_0 r
 \sin(\phi)\Bigg]_{k_0 =  \overline{\omega}(k)} \,.\label{shiftI2}
 \eea From (\ref{alplusbe},\ref{alminbe}) it is clear that the propagator in the medium features two
 Breit-Wigner complex poles corresponding to the two propagating
 modes in the medium.

In the expressions above we have only focused on the positive energy
modes. The expressions for the negative energy modes may be obtained
from the following relations which are consequences of the imaginary
parts of the self-energies and the dispersive representation valid
both for scalar and vector boson exchange (\ref{selfa}), \bea
\mathrm{Im}\mathds{A}(-k_0,k;\mu) &  = &
\mathrm{Im}\mathds{A}(k_0,k;-\mu)~~;~~
\mathrm{Re}\mathds{A}(-k_0,k;\mu)   =   -
\mathrm{Re}\mathds{A}(k_0,k;-\mu)~, \label{relA} \\
\mathrm{Im}\mathds{B}(-k_0,k;\mu) &  = & -
\mathrm{Im}\mathds{B}(k_0,k;-\mu)~~;~~
\mathrm{Re}\mathds{B}(-k_0,k;\mu)    =
\mathrm{Re}\mathds{B}(k_0,k;-\mu)\,. \label{relB} \eea These
properties can be read-off the explicit expressions for the
imaginary parts of the self-energies given in the appendix equations
(\ref{Imsigdec}-\ref{PI1}) for the standard model contributions and
equations (\ref{imsigbsm}-\ref{PI1bsm}) for the scalar exchange
contributions. The matrices $\mathds{A}$ are extracted from the
coefficient of $\gamma^0$ and $\mathds{B}$ from the coefficients of
$\vec{\gamma}\cdot  \hat{k}$ in the self-energies respectively. The
relations for the real parts follow from the dispersive
representation (\ref{selfa}).

 In what follows we use the ultrarelativistic approximation \be
 \overline{\omega} (k) \simeq k + \frac{\overline{M}^{\,2}}{2k}~.  \label{URapx}\ee
 In the limit of interest $k/T \lesssim  1$ with  $M_1 \ll M_2 \sim M_s \sim
 \mathcal{O}(keV)$, the region $k < T \sim \mathcal{O}(100\,GeV)$ corresponds to a wide
 window in which the ultrarelativistic approximation is reliable.

 We note that the difference in the real part of the pole position in the ultrarelativistic
 limit becomes \be \Omega_2(k) - {\Omega}_1(k) \simeq \frac{\delta M^2}{2k} \rho_0 r \cos(\phi)\,.
 \label{freqdiff}\ee  From the expression (\ref{alfaangle})   for $|\tilde{\gamma}| > 1$ it
 follows that  when an  MSW resonance occurs, namely for $\theta_m = \pi/4$  resulting in  $\cos(\phi) =0$
 and the real part of the poles become \emph{degenerate}. This is
 in striking contrast with the quantum mechanical description of mixed neutrinos where no level
 crossing (or complete degeneracy) can occur. Indeed the degeneracy is a consequence of the fact
 that the self-energy is complex and only occurs when damping is strong in the sense that $|\tilde{\gamma} | >1$.

 The   degeneracy   near an MSW resonance for strong damping will
 necessarily result
 in a breakdown of adiabaticity during cosmological evolution. We
 analyze below the conditions required for this phenomenon to occur.

 Furthermore, as discussed in refs.\cite{kusepetra,boysnudm}
 decoupling and freeze-out of sterile neutrinos of neutrinos
 produced via scalar decay occurs near the electroweak scale, and it
 will be seen consistently that vector boson decay yields a production
 rate with a similar structure as for scalar decay therefore a similar
 range of temperatures in which sterile neutrino production by this mechanism is effective.

 Perturbation theory is reliable when the change in the dispersion
 relations (positions of the poles in the propagators) is  small. In the relativistic
 limit the (bare) poles correspond to $k_0 =k$ (for positive energy particles), therefore
 perturbation theory is valid for $ k \gg (\Omega_{1,2}-k) ~;~  \Gamma_{1,2}$  namely  $ k \gg  \Sigma(k,k) $ where $\Sigma$ is any of the self-energies. In the next  section     we obtain explicitly the self energies and in section (\ref{subsec:PT})  we
assess the regime of validity of the perturbative expansion.

 \subsection{Helicity dependence: right-handed sterile neutrinos and standard model interactions}

 We have purposely kept the general form of the self-energies and
 propagators in terms of the helicity projections $h=\pm1$. In the
 non-interacting massless case,   positive   energy left-handed
 particles correspond to  $h=-1$ and negative energy left-handed correspond to
 $h=1$, with the opposite assignment for right-handed particles. For
 the massive but ultrarelativistic case the mass term yields
 corrections to the handedness-helicity assignment of
 $\mathcal{O}(M^2/k^2)$.

 \medskip

 \underline{$\mathbf{h = -1}$: }  Neglecting subleading terms of $\mathcal{O}(\overline{M}^{\,2}/k^2)$ that multiply
  (bsm) right-handed contributions in the ultrarelativistic limit, we obtain

 \bea S(k) & = &    2k \Big[
\big(\mathds{A}_L-\mathds{B}_L \big)_{aa} +
\big(\mathds{A}_L-\mathds{B}_L \big)_{ss}  \Big]~, \label{S1hm1} \\
\Delta(k) & = & \frac{2 k }{\delta M^2}\Big[
\big(\mathds{A}_L-\mathds{B}_L \big)_{aa} -
\big(\mathds{A}_L-\mathds{B}_L \big)_{ss} \Big]~. \label{Del1hm1}\eea

\underline{$\mathbf{h= 1}$:} In this case the corrections of $\mathcal{O}(\overline{M}^{\,2}/k^2)$ multiply (sm)
left-handed contributions, which may be of the same order of the (bsm) right-handed contributions.
We find,

\bea S(k)   &  =  &  2k\Big[ \big(\mathds{A}_R -\mathds{B}_R
\big)_{aa} + \big(\mathds{A}_R-\mathds{B}_R \big)_{ss}     +
 \frac{\overline{M}^{\,2}}{4k^2}  \big(\mathds{A}_L+ \mathds{B}_L \big)_{aa} \Big]~, \label{S2hp1} \\
\Delta(k) & = & \frac{2k}{\delta M^2 }\Big[ \big(\mathds{A}_R-
\mathds{B}_R \big)_{aa} - \big(\mathds{A}_R- \mathds{B}_R \big)_{ss}
      + \frac{\overline{M}^{\,2}}{4k^2}   \big(\mathds{A}_L+
\mathds{B}_L \big)_{aa}  \Big]~. \label{Del2hp1} \eea

The terms proportional to $\overline{M}^{\,2}/4k^2$ only receive
contribution from the \emph{standard model} self-energies, whereas
the right-handed components only originate in the contributions
beyond the standard model which are suppressed by much smaller
Yukawa couplings. However the last contribution in (\ref{Del2hp1})
from (sm) interactions \emph{may be } of the same order as the (bsm)
contributions for a relevant range of $k$.   To see this note that
$\mathds{A}_R,\mathds{B}_R \sim Y^2_2 \sim 10^{-16}$, whereas
$\mathds{A}_L,\mathds{B}_L \sim g^2 \sim 0.4$ therefore with
$\overline{M} \sim \mathrm{KeV}$ and $k \lesssim 100\,\mathrm{GeV}$,
 it is clear that both contributions (bsm) and (sm) are of the same
order.

 The point of maintaining the helicity dependence throughout is that
 for the case of sterile neutrinos, namely the propagating modes
 ``2'' in the medium, the exchange of standard model vector bosons
 yields a contribution to the positive helicity and positive energy
 components, namely the right-handed component, which could be of
 the same order of the (bsm) contributions for small $k$ which is a region of interest for
 sterile neutrino production.

 \section{Real parts: mixing angles and MSW
 resonances}\label{sec:realparts}
 The dispersion relations (real parts of the poles) and the mixing angles in the medium are
 determined by the real parts of the self-energy, namely the ``index of refraction''. Whereas
 the neutral current tadpole contribution (\ref{sigta}) is real, the real part of the other
 contributions is obtained from the dispersive form (\ref{selfa}), namely
 \be \mathrm{Re} \Sigma(k_0,k) = \frac{1}{\pi}\int_{-\infty}^\infty d\omega ~\mathcal{P} \Bigg(
\; \frac{\mathrm{Im}\Sigma(\omega,\vec{k}\,)}{\omega-k_0   } \Bigg)\;
 .\label{Reselfa}\ee In general the real part  must be obtained numerically and is a function
 of three parameters $k_0,k,\mu$ which makes its exploration a daunting numerical task. However, progress
 can be made by focusing   on the ``on-shell'' contribution, namely setting $k_0\simeq k$, and
 neglecting the   dependence on $\mu$, which is warranted in the whole region of $k,T$ of interest,
 but for   $k/T,|\mu|/T \ll M/T$,  in which case  we provide below an accurate approximate form.

 In obtaining the
 real parts we consider only the finite temperature contribution, because the zero temperature part is
 absorbed in the renormalization of the parameters in the
 Lagrangian.

 \medskip

 \underline{\textbf{Scalars (bsm):}} For the real part of the scalar (bsm) self-energy
 we find for $k_0=k~; ~\mu=0$ \be \mathrm{Re}\Sigma_{bsm}(k,k) =
 \frac{Y^2 T}{16\pi^2} \Bigg\{ \gamma^0
 \Bigg[Af\bigg(\frac{k}{T};\frac{M}{T}\bigg)+Ab\bigg(\frac{k}{T};\frac{M}{T}\bigg)\Bigg]
 -\vec{\gamma}\cdot\uvk\Bigg[Bf\bigg(\frac{k}{T};\frac{M}{T}\bigg)+Bb\bigg(\frac{k}{T};
 \frac{M}{T}\bigg)\Bigg]\Bigg\}~,
 \label{ReSigbsm}\ee where $Af;Bf$ and  $Ab;Bb$  are  the fermionic and  bosonic
 contributions respectively and $Y = Y_{1,2}$ for $\sigma,\varphi$ exchange. Figs. (\ref{fig:coeffuncs}) show $Af;Bf$ and  $Ab;Bb$
 for $M/T = 1,2,3$ as a function of $k/T$.

\begin{figure}[h]
\begin{center}
\includegraphics[height=5cm,width=8cm]{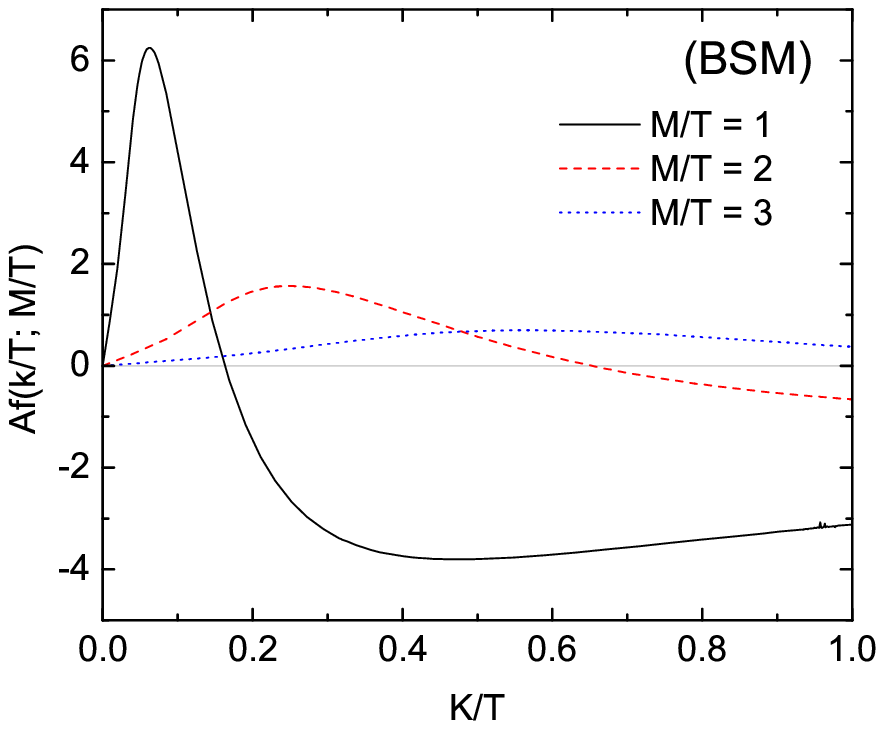}
\includegraphics[height=5cm,width=8cm]{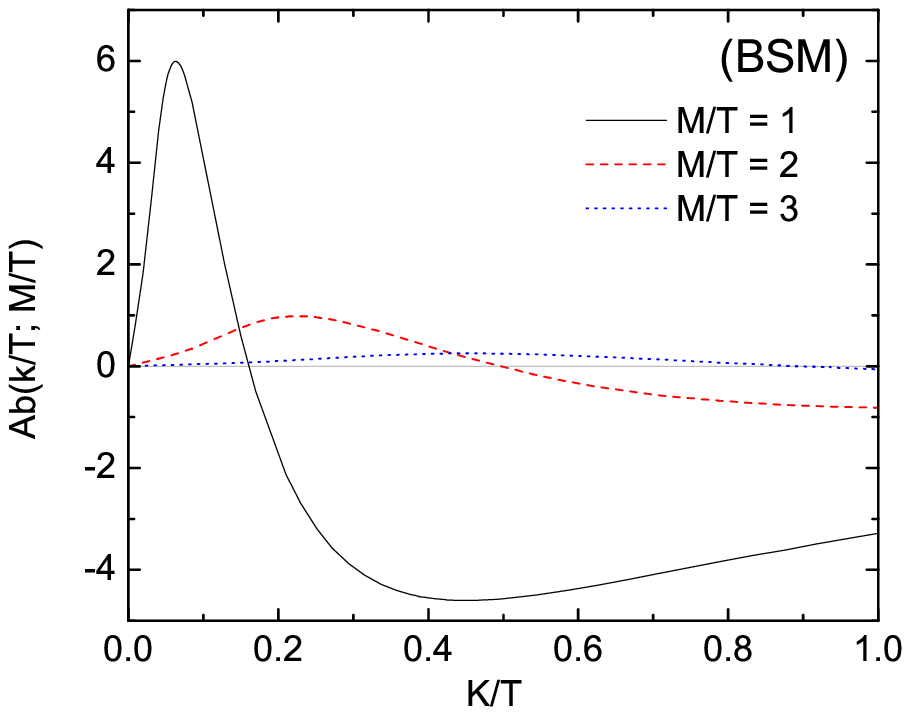}
\includegraphics[height=5cm,width=8cm]{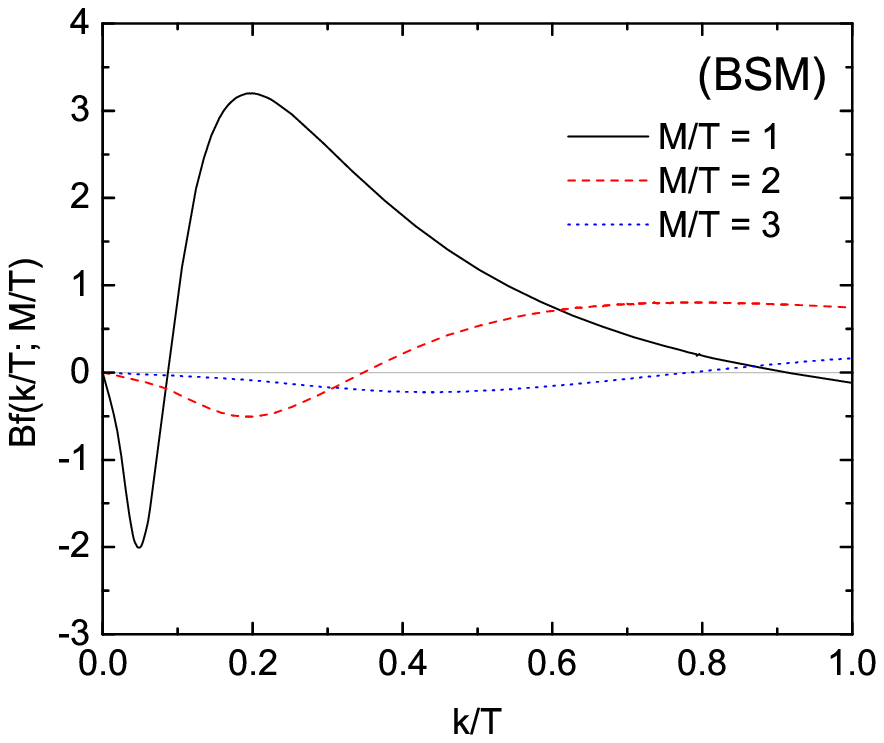}
\includegraphics[height=5cm,width=8cm]{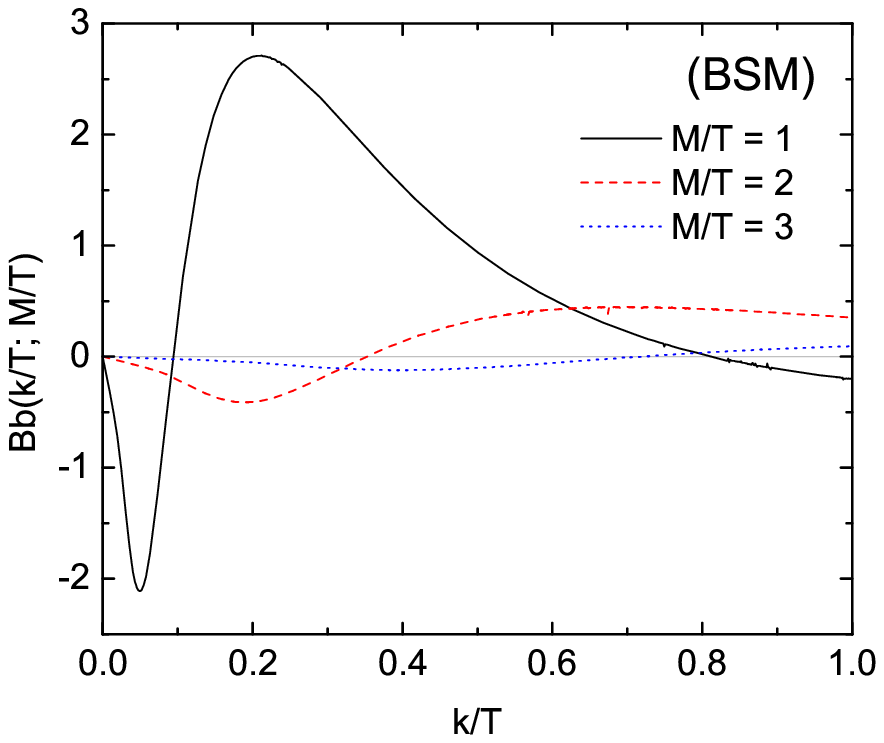}
\caption{ The functions $Af(k/T;M/T); Ab(k/T;M/T);
Bf(k/T;M/T);Bb(k/T;M/T)$ as a function of $k/T$ for
 $M/T = 1,2,3$} for $\mathrm{Re}\Sigma_{bsm}$.\label{fig:coeffuncs}
\end{center}
\end{figure}

For $\Sigma^{(2)}_{\sigma,\varphi}$ the intermediate fermion line
corresponds to a sterile-like neutrino, therefore for these
contributions we must set $Af=0;Bf=0$, under the assumption that the
sterile neutrino population can be neglected and the propagator for
the internal line is the vacuum one. For the mixing angle the
relevant contribution is $A-B$. Figs. (\ref{fig:coefdiffuncs})
display $Af-Bf$ and $Ab-Bb$ for $M/T =1,2,3$ as a function of $k/T$.

\begin{figure}[h]
\begin{center}
\includegraphics[height=5cm,width=8cm]{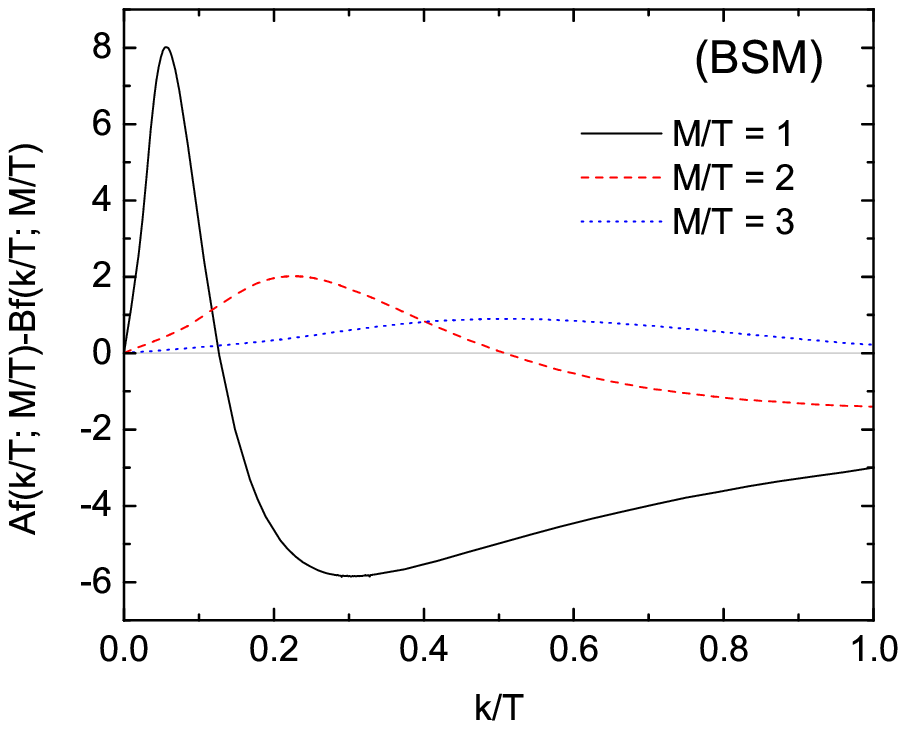}
\includegraphics[height=5cm,width=8cm]{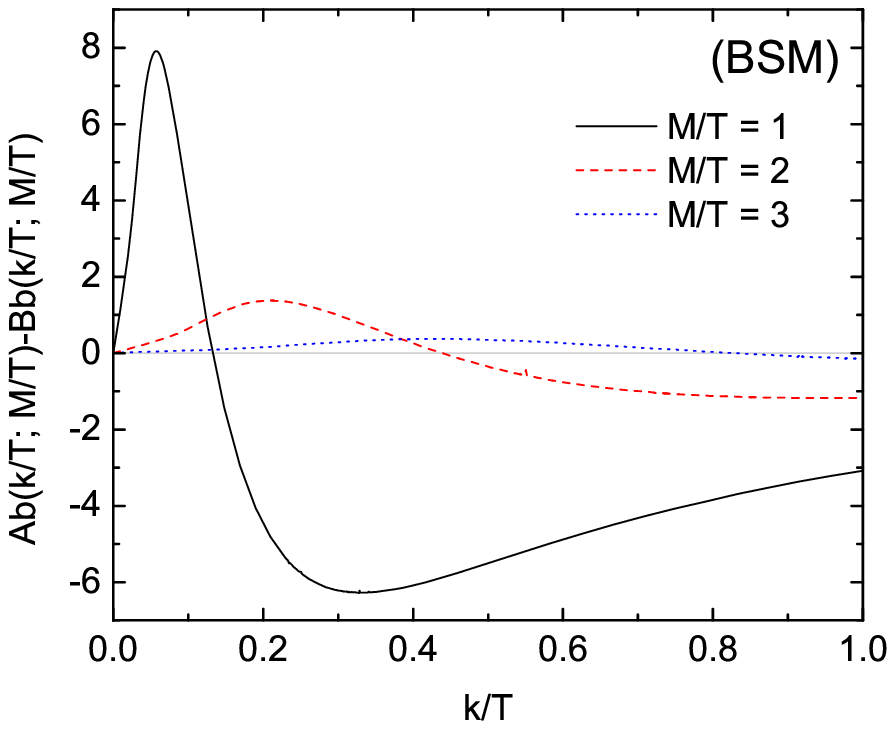}
\caption{ The functions $Af(k/T;M/T)- Bf(k/T;M/T); Ab(k/T;M/T)-
Bb(k/T;M/T)$ as a function of $k/T$ for
 $M/T = 1,2,3$}  for the case of scalars (bsm). \label{fig:coefdiffuncs}
\end{center}
\end{figure}

We note that the fermionic and bosonic contributions $Af,Ab$ are
qualitatively very similar and the same property holds for $Bf,Bb$.
Therefore neglecting the fermionic contributions both for
$\Sigma^{(1)}$ does \emph{not} affect the results and the
conclusions in a substantial manner.

This observation confirms that the general results presented below
are robust \emph{even} when the neutrinos ``1'' are not thermalized
and their propagators are the vacuum ones.

\medskip

Although an analytic form for the full range of $k_0;k;\mu$ is not
available, we obtain an analytic expression for the relevant case
$k_0/T, k/T, \mu/T \ll M/T \sim 1 $. We find to leading order in the
small ratios $k_0/T;k/T;\xi = \mu/T$,

\be \mathrm{Re}\Sigma^{(1)}_{\sigma}(k_0,k) =  \frac{Y^2_1 T^2}{
M^2_\sigma} \Bigg\{\gamma^0 \Bigg[-\frac{T\,\xi}{12} \Bigg(1 +
\frac{\xi^2}{\pi^2} \Bigg)+
  \frac{7\pi^2}{120}
  \frac{ k_0\,T^2}{M^2_\sigma} \Big[1+ F[M_\sigma/T]\Big] \Bigg] -\vec{\gamma}\cdot \hat{k} \Bigg[- \frac{7\pi^2}{360}
  \frac{ k \,T^2}{M^2_\sigma} \Big[1+ J[M_\sigma/T]\Big] \Bigg]   \Bigg\}~, \label{sig1sig}\ee

\be \mathrm{Re}\Sigma^{(1)}_{\varphi}(k_0,k) =  \frac{Y^2_2 T^2}{
M^2_\varphi} \Bigg\{\gamma^0 \Bigg[-\frac{T\,\xi}{12} \Bigg(1 +
\frac{\xi^2}{\pi^2} \Bigg)+
  \frac{7\pi^2}{120}
  \frac{ k_0\,T^2}{M^2_\varphi} \Big[1+ F[M_\varphi/T]\Big] \Bigg] -\vec{\gamma}\cdot \hat{k} \Bigg[- \frac{7\pi^2}{360}
  \frac{ k \,T^2}{M^2_\varphi} \Big[1+ J[M_\varphi/T]\Big] \Bigg]   \Bigg\}~, \label{sig1varfi}\ee

  \be \mathrm{Re}\Sigma^{(2)}_{\sigma}(k_0,k) =  \frac{Y^2_1 T^2}{  M^2_\sigma}
\Bigg\{\gamma^0 \Bigg[
  \frac{7\pi^2}{120}
  \frac{ k_0\,T^2}{M^2_\sigma}   F[M_\sigma/T]\Big]  -\vec{\gamma}\cdot \hat{k} \Bigg[- \frac{7\pi^2}{360}
  \frac{ k \,T^2}{M^2_\sigma}   J[M_\sigma/T]  \Bigg]   \Bigg\}~, \label{sig2sig}\ee

\be \mathrm{Re}\Sigma^{(2)}_{\varphi}(k_0,k) =  \frac{Y^2_2 T^2}{
M^2_\varphi} \Bigg\{\gamma^0 \Bigg[
  \frac{7\pi^2}{120}
  \frac{ k_0\,T^2}{M^2_\varphi}   F[M_\varphi/T]  \Bigg] -\vec{\gamma}\cdot \hat{k} \Bigg[- \frac{7\pi^2}{360}
  \frac{ k \,T^2}{M^2_\varphi}   J[M_\varphi/T]  \Bigg]   \Bigg\}~,
  \label{sig2varfi}\ee where

  \be J(m)    =    \frac{120}{7\pi^4}
\int_0^\infty dq ~\frac{q^2}{W_q} N_B(W_q) \Big[W^2_q +
\frac{m^2}{2}\Big]  ~~;~~F(m)    =    \frac{120}{7\pi^4}
\int_0^\infty dq~\frac{q^2}{W_q} N_B(W_q) \Big[W^2_q -
\frac{m^2}{2}\Big]  \,. \label{fjofm}\ee

These functions are displayed in Fig. (\ref{fig:functions}), they
are $\mathcal{O}(1)$ in the region of interest
$M_{\sigma,\varphi}\sim T$.

A comprehensive numerical study of $Af,Ab,Bf,Bb$ confirms the
validity of the above approximations for $k_0=k,\mu=0$ for $k/T \ll
1$.

\medskip

\underline{\textbf{Vector bosons (sm):}}  Similarly, for the real part of the
(sm) self-energy
 we find for $k_0=k~~;\mu=0$ \be \mathrm{Re}\Sigma_{sm}(k,k) =
 \frac{g^2_{sm} T}{16\pi^2} \Bigg\{ \gamma^0
 \Bigg[Af\bigg(\frac{k}{T};\frac{M}{T}\bigg)+Ab\bigg(\frac{k}{T};\frac{M}{T}\bigg)\Bigg]
 -\vec{\gamma}\cdot\uvk\Bigg[Bf\bigg(\frac{k}{T};\frac{M}{T}\bigg)+Bb\bigg(\frac{k}{T};\frac{M}{T}\bigg)\Bigg]~,
 \label{ReSigsm}\ee where we use the same definition, namely $Af;Bf$ and  $Ab;Bb$  are  the fermionic and  bosonic
 contributions respectively. Figs. (\ref{fig:coeffuncssm}) show $Af;Bf$ and
 $Ab;Bb$ and Fig.(\ref{fig:coefdiffuncssm}) shows $Af(k/T;M/T)- Bf(k/T;M/T); Ab(k/T;M/T)-
Bf(k/T;M/T)$  for $M/T = 1,2,3$ as a function of $k/T$.

\begin{figure}[h]
\begin{center}
\includegraphics[height=5cm,width=8cm]{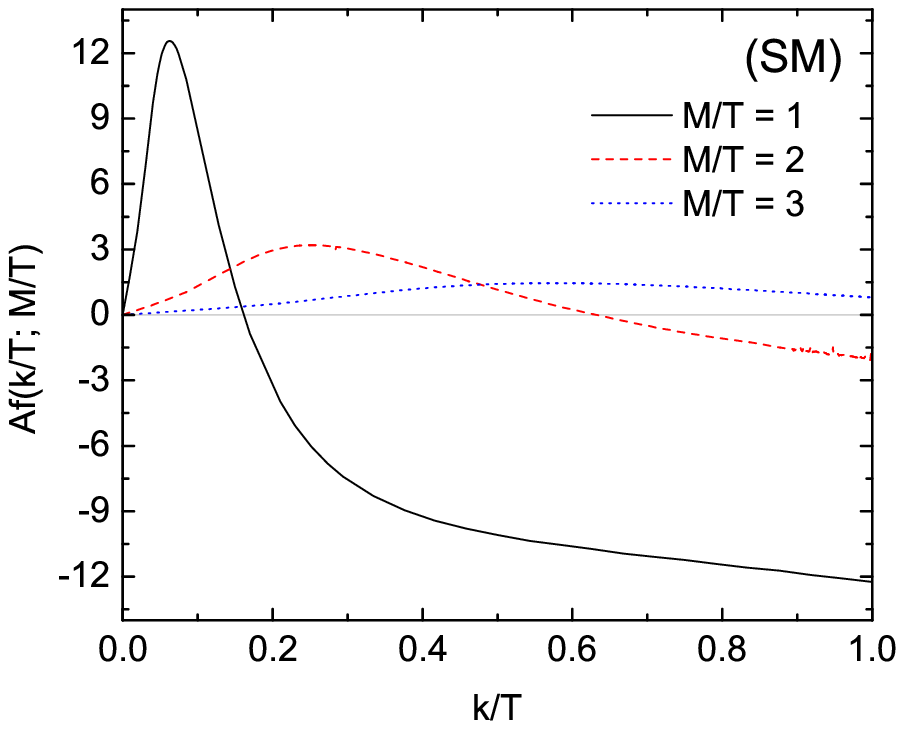}
\includegraphics[height=5cm,width=8cm]{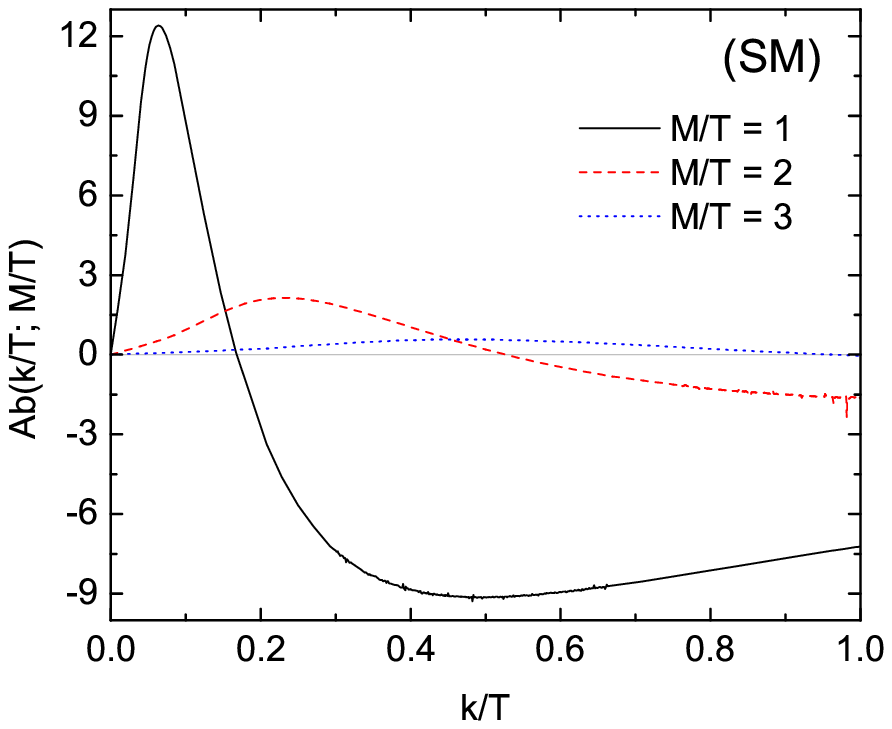}
\includegraphics[height=5cm,width=8cm]{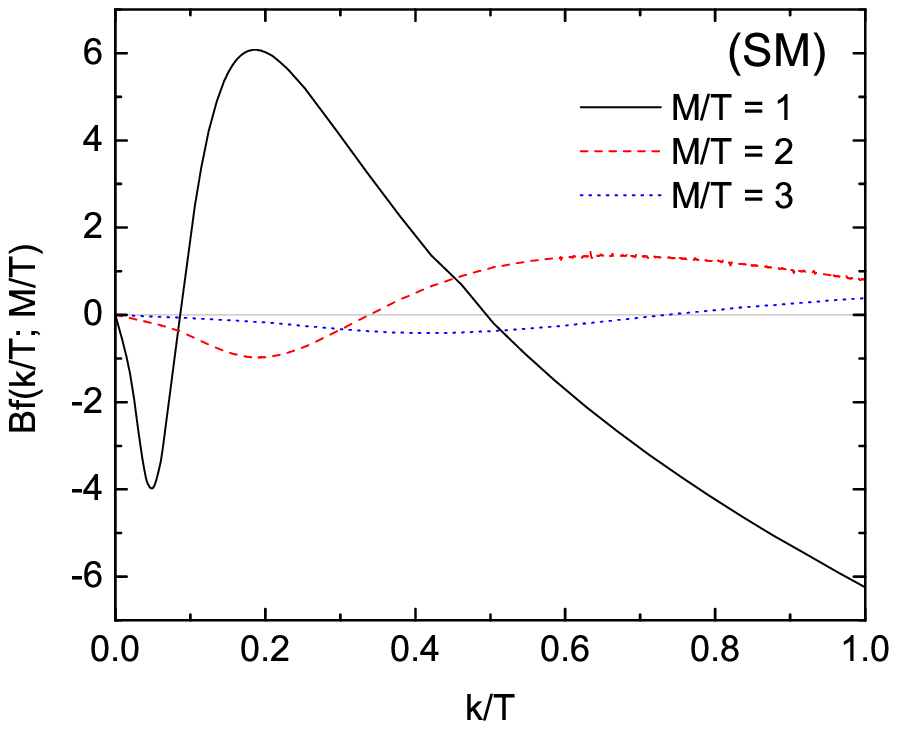}
\includegraphics[height=5cm,width=8cm]{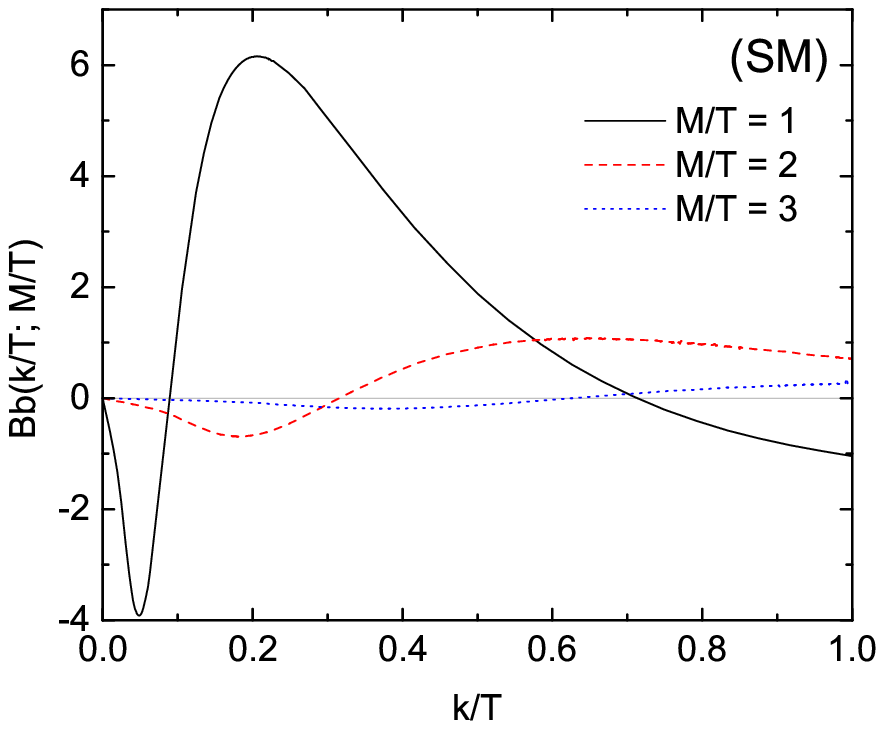}
\caption{ The functions $Af(k/T;M/T); Ab(k/T;M/T);
Bf(k/T;M/T);Bb(k/T;M/T)$ as a function of $k/T$ for
 $M/T = 1,2,3$} for the standard model contributions (sm) with $\mu=0$.\label{fig:coeffuncssm}
\end{center}
\end{figure}

Just as in the (bsm) case analyzed above, we note that the fermionic and bosonic contributions
$Af;Ab$ are qualitatively similar and the same holds for $Bf;Bb$. Again this observation confirms that
our results are robust, independently of whether \emph{any} of the neutrino modes is thermalized.

\begin{figure}[h]
\begin{center}
\includegraphics[height=5cm,width=8cm]{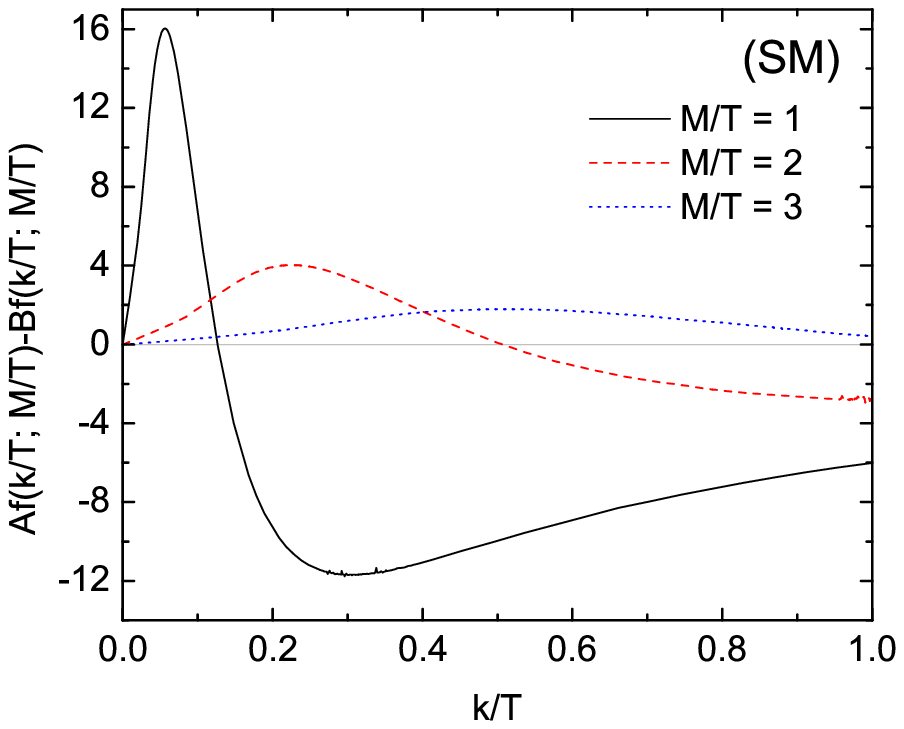}
\includegraphics[height=5cm,width=8cm]{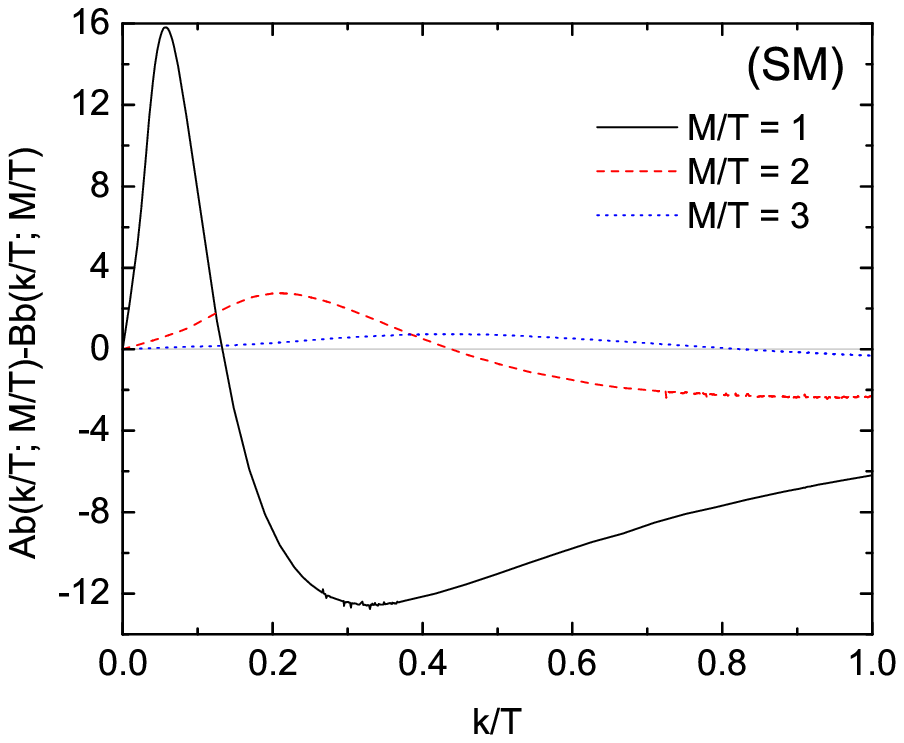}
\caption{ The functions $Af(k/T;M/T)- Bf(k/T;M/T); Ab(k/T;M/T)-
Bb(k/T;M/T)$ as a function of $k/T$ for
 $M/T = 1,2,3$}  for the standard model contributions (sm) with $\mu=0$. \label{fig:coefdiffuncssm}
\end{center}
\end{figure}

We also obtain the analytic forms for
$\mathrm{Re}\Sigma_{sm}(k_0;k)$ for $k_0/T,k/T,\mu/T \ll M_{W,Z}/T
\sim 1$. To leading order in these small ratios we find

 \be
 \mathrm{Re}\Sigma^{(1)}_{nc}(k_0,k) =  \frac{g^2 T^2}{4 M^2_W} \Bigg\{\gamma^0 \Bigg[-\frac{T\,\xi}{4} \Bigg(1 + \frac{\xi^2}{\pi^2} \Bigg)+
  \frac{7\pi^2}{60}
  \frac{ k_0\,T^2}{M^2_Z} \Big[1+ G[M_Z/T]\Big] \Bigg] -\vec{\gamma}\cdot \hat{k} \Bigg[- \frac{7\pi^2}{180}
  \frac{ k \,T^2}{M^2_Z} \Big[1+ G[M_Z/T]\Big] \Bigg]   \Bigg\}~, \label{sig1nc}\ee

  \be \mathrm{Re}\Sigma^{(2)}_{nc}(k_0,k) =  \frac{g^2 T^2}{4 M^2_W} \Bigg\{\gamma^0
  \Bigg[
  \frac{7\pi^2}{60}
  \frac{ k_0\,T^2}{M^2_Z}   G[M_Z/T]  \Bigg] -\vec{\gamma}\cdot \hat{k} \Bigg[- \frac{7\pi^2}{180}
  \frac{ k \,T^2}{M^2_Z}   G[M_Z/T]  \Bigg]   \Bigg\}~, \label{sig2nc}\ee

   \be \mathrm{Re}\Sigma_{cc,sm}(k_0,k) =  \frac{g^2 T^2}{2 M^2_W} \Bigg\{\gamma^0
   \Bigg[
  \frac{7\pi^2}{60}
  \frac{ k_0\,T^2}{M^2_W} \Big[1+ G[M_W/T]\Big] \Bigg] -\vec{\gamma}\cdot \hat{k} \Bigg[- \frac{7\pi^2}{180}
  \frac{ k \,T^2}{M^2_W} \Big[1+ G[M_W/T]\Big] \Bigg]   \Bigg\}~. \label{sigcc}\ee

In the charged current contribution we have neglected the asymmetry
of the charged lepton because it is of the order of the baryon
asymmetry. In the above expressions \be G[m] = \frac{120}{7\pi^4}
\int_0^\infty dq~\frac{q^2}{W_q} N_B(W_q) \Big[W^2_q -
\frac{m^2}{4}\Big]  ~~;~~N_B(W_q) =  \frac{1}{e^{W_q}-1}~~;~~
W_q=\sqrt{q^2+m^2}\,. \label{fofm}\ee This function is depicted in
Fig.(\ref{fig:functions}), it is $\mathcal{O}(1)$ in the region of
interest $T \sim M_{Z,W}$.

\begin{figure}[h]
\begin{center}
\includegraphics[height=5cm,width=7cm]{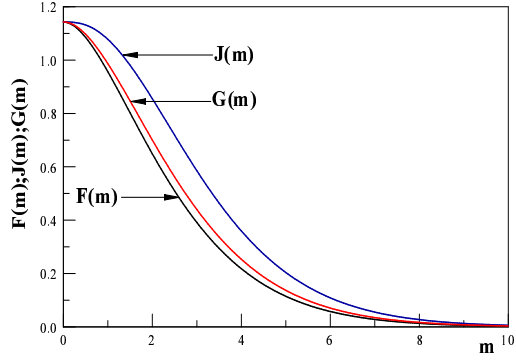}
\caption{The functions $F(m);J(m);G(m)$ vs $m=M/T$.}
\label{fig:functions}
\end{center}
\end{figure}

The validity of these approximations for $k_0=k,\mu=0$ is confirmed
by the numerical analysis of $Af,Ab,Bf,Bb$ for $k/T \ll 1$.

It is remarkable that the leading order in $k_0/T,k/T$ but for
$M_{W,Z} \sim T$ reproduce the results of
references\cite{notzold,boyhonueu} which were obtained in the low
energy limit $T,\mu \ll M_{W,Z}$. The numerical analysis carried out
for $k_0=k;\mu=0$ confirms that for $M/T \gg 1$ the range of
validity of the lowest order approximation in $k/T$ increases and
merges with the results given above in eqns.
(\ref{sig1sig}-\ref{sigcc}) up to $k/T \sim 1$.

\subsection{Mixing angles and MSW resonances:}
As shown in detail in the previous sections, the mixing angle in the medium $\theta_m$ determined
by the relations (\ref{mediumangle}) depends on $k_0,k$ and the helicity $h$. On the mass shell
of the propagating modes in the medium we can replace $k_0   \simeq  k$ in the expressions for the real
part of the matrices $\mathrm{Re}\mathds{A}~;~\mathrm{Re}\mathds{B}$ for  $\Delta_R$, namely
  the real part of eqns. (\ref{Del1hm1},\ref{Del2hp1}) for $h=\mp 1$ respectively. For $\mu=0;k_0 =k$ and
general $k,M$ the fermionic and bosonic contributions to the real parts of the  (bsm) self-energies are given by eqn. (\ref{ReSigbsm}) where the fermionic ($Af,Bf$) and bosonic ($Ab,Bb$) contributions
are depicted in figs. (\ref{fig:coeffuncs}-\ref{fig:coefdiffuncs}). The real parts of the (sm) self-energies are given by eqn. (\ref{ReSigsm}) and the fermionic and bosonic contributions depicted in figs.(\ref{fig:coeffuncssm}-\ref{fig:coefdiffuncssm}).

These figures distinctly show that the contributions $Af,Ab$ and
$Bf,Bb$ for (bsm) and (sm) self-energies are qualitatively the same,
with only a quantitative difference in the amplitudes. A remarkable
result is that these functions \emph{change sign}. In particular the
combinations $Af-Bf,Ab-Bb$ which enter in $\Delta_R$ change sign at
a value of $k/T$ that depends on the ratio $M/T$. For $M/T \sim 1$
these differences vanish  at $k/T \simeq 0.2$.  A numerical
exploration reveals that the sign change persists until $M/T \simeq
3 $ but occurs at monotonically larger values of $k/T$. This
behavior is shown in the figures above. We find that for $M/T
\gtrsim 3$ the change in sign occurs for $k>>T$ or does not occur at
all. On the mass shell $k_0 \sim k$ and for $\mu=0$ this study
reveals that $\Delta_R$ is \emph{negative} in a wide region of
momentum for $M/T \lesssim 1$. This fact entails that there are MSW
resonances near the momentum regions where the coefficient functions
change sign, even in \emph{absence of a lepton asymmetry}.
 To understand this important point more clearly let us study the  case $h=\mp 1$ separately.

 \medskip

 \underline{$\mathbf{h=-1}$:} In this case
 $\Delta$ is given by eqn. (\ref{Del1hm1}), furthermore from eqn. (\ref{sigmass}) it follows that
  $(\mathds{A}_L-\mathds{B}_L)_{ss}$ is determined by the (bsm) contributions which are suppressed by
   small Yukawa couplings $Y \lesssim 10^{-8}$ as compared to the (sm) contributions. Therefore the
   (bsm) contribution can be neglected and $\Delta_R$
is determined by the (sm) contributions given by eqns. (\ref{ReSigsm}), furthermore approximating
 $\cos(\theta) \sim 1;\sin (\theta)\sim 0$ in  eqn. (\ref{sigmaaa})
 and $\delta M^2 \simeq M^2_s$,
 we find (for $\mu=0;h=-1$) \bea \Delta_R(k) & \simeq &  \frac{g^2 }{16\pi^2
 }\bigg(\frac{T }{M_s}\bigg)^2\bigg(\frac{k}{T}\bigg)
 \Bigg\{\Big[ Af\Big(\frac{k}{T},\frac{M_W}{T}\Big)+ Ab\Big(\frac{k}{T},\frac{M_W}{T}\Big)-
 Bf \Big(\frac{k}{T},\frac{M_W}{T}\Big)- Bb \Big(\frac{k}{T},\frac{M_W}{T}\Big) \Big] \nonumber \\
  & + &  \frac{1}{2\cos(\theta_w)}\Big[ Af\Big(\frac{k}{T},\frac{M_Z}{T}\Big)+ Ab\Big(\frac{k}{T},
  \frac{M_Z}{T}\Big)-Bf \Big(\frac{k}{T},\frac{M_Z}{T}\Big)- Bb \Big(\frac{k}{T},\frac{M_Z}{T}\Big) \Big]
    \Bigg\}~.  \label{DelRhm1ap} \eea

    Taking as representative $T \sim 100\,\mathrm{GeV};M_s \sim
    \mathrm{keV}$ it follows that \be \frac{g^2 }{16\pi^2
 }\bigg(\frac{T }{M_s}\bigg)^2 \simeq 2.7 \times 10^{13}\,.
 \label{coupe}\ee Figures (\ref{fig:coefdiffuncssm}) show that for
 $M_{W,Z}/T \lesssim 3$ there is a region in $k/T$ in which the bracket in
 (\ref{DelRhm1ap}) is \emph{negative} and there is a value
 $(k/T)_c$ that increases with  $M/T$ at which the bracket vanishes, for example from the Fig. (\ref{fig:coefdiffuncssm}) we find
 $(k/T)_c \sim 0.2; 0.45;1$ for $M/T \sim 1,2,3$ respectively. For $k/T < (k/T)_c$ the
 bracket is positive (for $\mu=0$) whereas for $k/T > (k/T)_c$
it is \emph{negative}, therefore there is a value of $k/T$ at which
the resonance condition (\ref{MSWcond}) is fulfilled. Since the
coefficient of the bracket is $\approx 10^{13}$ (eqn. (\ref{coupe}))
and the terms inside the bracket are of $\mathcal{O}(1)$ for $k/T
\lesssim 1$, and $\cos(\theta) \sim 1$ it follows that the MSW
resonance occurs for a value of $k/T$ such that the bracket $\sim
10^{-13}$ namely for $k/T \sim (k/T)_c$. The large coefficient
(\ref{coupe}) results in a \emph{very narrow} MSW resonance as can
be seen as follows, expanding $\Delta_R$ near $(k/T)_c$ as \be
\Delta_R(k) \simeq  -\kappa \Bigg( \Big( \frac{k}{T}\Big) - \Big(
\frac{k}{T}\Big)_c \Bigg) +\cdots ~~;~~ \kappa > 0~,
\label{delexpa}\ee where
 $\kappa \gtrsim 10^{13}$ for $ M_{W,Z}/T\lesssim 3$ (see Fig. (\ref{fig:coefdiffuncssm})) and approximating
 $\cos(2\theta)\sim 1$ we find \be \sin^2(2\theta_m) \simeq
\frac{\epsilon^2}{\Bigg[\Bigg( \Big( \frac{k}{T}\Big) - \Big(
\frac{k}{T}\Big)_c -\frac{1}{\kappa} \Bigg)^2+\epsilon^2\Bigg]}
~~;~~\epsilon= \sin(2\theta)/\kappa ~.\label{sharpmsw}\ee For example
taking $\sin(2\theta) \sim 10^{-5}$\cite{kuseobs} it follows that
$\epsilon \lesssim 10^{-18}$ which makes the resonance very narrow.
During cosmological expansion the ratio $M/T(t)$ increases with the
scale factor, while the ratio $k/T$ (with $k$ the physical momentum)
is fixed. Therefore, for a fixed value of
 $k/T < 1 $ as $M/T$ increases
  the resonance is crossed very sharply.

  \medskip

\underline{$\mathbf{h=1}$:}
To assess the possibility of MSW resonances for  $h=1$ we need the real part of (\ref{Del2hp1}). From
(\ref{sigmaaa}) and (\ref{ReSigbsm}) it follows that $(\mathds{A}_R-\mathds{B}_R)_{aa} \propto Y^2_1~;~
(\mathds{A}_R-\mathds{B}_R)_{ss} \propto Y^2_2$, since $Y_2 \gg Y_1$ we can neglect the first term (corresponding to
$\sigma$-exchange). Similarly in the term $(\mathds{A}_R-\mathds{B}_R)_{ss}$  we neglect
the contribution from $\sigma$-exchange and approximate $\cos(\theta) \sim 1~;~ \sin(\theta) \sim 0$ in (\ref{sigmass}),
 hence only $\Sigma^{(2)}_\varphi$ contributes to $\Sigma_{ss}$. Furthermore, approximating $\delta M^2 \sim \overline{M}^{\,2} \sim M_s^2$ we finally find for $h=1~;~\mu=0$,
\bea \Delta_R(k) \simeq && -\Bigg(\frac{Y_2 T}{\sqrt{8}\pi M_s}\Bigg)^2 \Bigg(\frac{k}{T}\Bigg)
\Bigg[Ab\Big(\frac{k}{T},\frac{M_\varphi}{T} \Big)-Bb\Big(\frac{k}{T},\frac{M_\varphi}{T} \Big) \Bigg]
\nonumber \\ && + \frac{g^2 }{128\pi^2
 }\bigg(\frac{T }{k}\bigg)
 \Bigg\{\Big[ Af\Big(\frac{k}{T},\frac{M_W}{T}\Big)+ Ab\Big(\frac{k}{T},\frac{M_W}{T}\Big)+
 Bf \Big(\frac{k}{T},\frac{M_W}{T}\Big)+ Bb \Big(\frac{k}{T},\frac{M_W}{T}\Big) \Big] \nonumber \\
  &  & + \frac{1}{2\cos(\theta_w)}\Big[ Af\Big(\frac{k}{T},\frac{M_Z}{T}\Big)+ Ab\Big(\frac{k}{T},
  \frac{M_Z}{T}\Big)+Bf \Big(\frac{k}{T},\frac{M_Z}{T}\Big)+ Bb \Big(\frac{k}{T},\frac{M_Z}{T}\Big) \Big]
    \Bigg\}~,\label{num1} \eea where in the first line the $Ab;Bb$ are (bsm) displayed in figs. (\ref{fig:coeffuncs}).

    We note that with $T \sim 100 \mathrm{GeV}, M_s \sim \mathrm{KeV}$ the value of the $Y_2$ (see
    eqn. (\ref{Ys}) is such that $Y_2 T/M_s \sim \mathcal{O}(1)$, therefore fig (\ref{fig:coefdiffuncs}) (right panel)
    suggests that the (bsm) contribution may yield an MSW resonance in the region $k/T \lesssim 0.15~;~M_\varphi \sim T$,
    where the (bsm) contribution $Ab-Bb$ is \emph{positive} and large. Since
    $g^2/128\pi^2 \sim 3.4 \times 10^{-4}$ and $Af+Bf;Ab+Bb \sim \mathcal{O}(1)$ for $k/T \lesssim 1$ it
    follows that the (sm) contribution to $\Delta_R$ is \emph{subleading} for $k/T \lesssim 1$ and the (bsm)
    contribution \emph{may} lead to an MSW resonance in this region depending on the parameters of the
    extension (bsm).

    \medskip

    $\mathbf{ \mu \neq 0~;~k/T \ll M/T \sim 1 }:$

    \medskip

    The above results are valid for $\mu =0$, for $\mu\neq 0$ a full
    numerical evaluation of the real parts of the kernel is not
    available, however, the bounds on the lepton asymmetry from
    ref.\cite{steigman} suggest that $|\mu/T| \lesssim 0.02 \ll 1$
    and we can obtain a reliable understanding of the influence of
    the lepton asymmetry (in the neutrino sector) by focusing on the
    region of $k/T \ll 1$, in which we can use the results
    (\ref{sig1sig}-\ref{sig2varfi}) for (bsm) and (\ref{sigta}) along
    with  (\ref{sig1nc}-\ref{sigcc}) for (sm) and
    approximate
    $\cos(\theta) \sim 1;\sin(\theta)\sim 0$ in
    (\ref{sigmaaa},\ref{sigmass}) and $\delta M^2 \sim M^2_s$.

    For $h=-1$ again we neglect the (bsm) contributions to
    $\Delta_R(k)$ in (\ref{Del1hm1}), and   for $\mu/T;k/T \ll
    1$ we obtain, \be  \Delta_R(k) \simeq \frac{g^2 T^3 k}{M^2_W
    M^2_s}\Bigg\{ -\frac{5~\xi}{24}  + \frac{7\pi^2}{90}
    \Big(\frac{k}{T}\Big) \Bigg[\Big(\frac{T}{M_Z} \Big)^2
    \Bigg(1+G\Big(\frac{M_Z}{T} \Big)
    \Bigg)+2\Big(\frac{T}{M_W} \Big)^2
    \Bigg(1+G\Big(\frac{M_W}{T} \Big)
    \Bigg)\Bigg]\Bigg\} \,.\label{delRsmal}\ee We note that for $T
    \sim M_W; M_s\sim \mathrm{KeV}$ the prefactor \be \frac{g^2 T^3 k}{M^2_W
    M^2_s} \sim 10^{16}  \Big(\frac{k}{T}\Big) \label{prefc}\ee and
    the resonance condition (\ref{MSWcond}) can be fulfilled for $\xi > 0$ when the bracket in (\ref{delRsmal}) approximately
    vanishes, namely for \be \Big(\frac{k}{T}\Big) \sim
    \frac{25~\xi}{56\pi^2} \Rightarrow k \sim 0.05 ~\mu~, \label{kvalres}\ee where we have used
    $G(M_{W,Z}/T) \sim 1$ for $T\sim M_W$, a result that can be
    gleaned from Fig. (\ref{fig:functions}). For $\xi >0$ this MSW resonance occurs for
    antineutrinos (namely $k_0=-k$), a result that follows from the relations (\ref{relA},\ref{relB}).

 \medskip

    Similarly, for $h = 1$ and $\mu/T, k/T \ll 1$, we obtain
\begin{eqnarray}
\Delta_R(k) & \simeq & -\Big(\frac{Y_2^2 T^2}{M_s^2}\Big)\Big(\frac{T^4}{M_{\varphi}^4}\Big)\Big(\frac{k^2}{T^2}\Big)\frac{7\pi^2}{180}\left[J(\frac{M_{\varphi}}{T})+3F(\frac{M_{\varphi}}{T})\right] \nonumber \\
& & + \frac{g^2T^2}{M_W^2}\Big(\frac{T}{k}\Big)\Bigg\{-\frac{5 \xi}{192}+\frac{7\pi^2}{1440}\Big(\frac{k}{T}\Big)\Bigg[\Big(\frac{T}{M_Z}\Big)^2\Bigg(1+G\Big(\frac{M_Z}{T}\Big)\Bigg)+2\Big(\frac{T}{M_W}\Big)^2\Bigg(1+G\Big(\frac{M_W}{T}\Big)\Bigg)\Bigg]\Bigg\}. \nonumber \\
\label{delRsmal2}
\end{eqnarray}
Obviously, there is a competition between (sm) and (bsm) contributions in eqn.~(\ref{delRsmal2}).
 When $T\sim M_{W,Z,\varphi}$, $J(1), F(1), G(1)\sim 1$ and $(Y_2^2 T^2)/M_s^2 \sim 1$. Therefore,
 the (bsm) contribution to $\Delta_R(k)$ is
\begin{eqnarray}
\Delta_R^{(bsm)} \sim -\frac{7\pi^2}{45}\Big(\frac{k}{T}\Big)^2 = -1.54 \Big(\frac{k}{T}\Big)^2,
\end{eqnarray}
and the (sm) contribution to $\Delta_R(k)$ reads
\begin{eqnarray}
\Delta_R^{(sm)} \sim 0.1\Big(\frac{T}{k}\Big)\left[-\frac{5\xi}{192}+\Big(\frac{k}{T}\Big)\frac{7\pi^2}{240}\right]
\sim 0.029-3 \times 10^{-3} \Big(\frac{T}{k}\Big)~\xi.
\end{eqnarray}
The resonance happens for $\Delta_R(k)\sim -1$, namely
\begin{eqnarray}
 3 \times 10^{-3}
\Big(\frac{T}{k}\Big)~\xi \sim 1.029. \label{smallexp}
\end{eqnarray}
Obviously, one is always able to find a value of $k/T$ to satisfy eqn.~(\ref{smallexp}) for any given positive lepton
asymmetry $\xi$. For $|\xi| \sim 10^{-2}$ consistent with the WMAP and BBN data\cite{steigman}, we obtain
\begin{eqnarray}
\frac{k}{T}   \sim 3\times 10^{-3}~ \xi \sim 3\times 10^{-5}\,.
\end{eqnarray}
Note that the asymmetry term from (sm) contribution dominates over the (bsm) contribution,
which is different from $\mu = 0$ case where (bsm) contribution would dominate as shown in (\ref{num1}).
    This analysis leads us to conclude that for a lepton asymmetry
    hidden in the neutrino sector compatible with the bounds from
    ref.\cite{steigman} there is the possibility of \emph{two} MSW
    resonances.

 \section{Imaginary parts: widths from vector and scalar boson decay.  }

 The quasiparticle widths $\Gamma_{1,2}(k)$  are given by eqns. (\ref{shiftI1}),(\ref{shiftI2}). Analyzing the
 explicit expressions for the imaginary parts of the (sm) and (bsm)
 contributions given in the appendix, equations
 \ref{Imsigdec}-\ref{PI1}, and  \ref{imsigbsm}-\ref{PI1bsm}
 respectively, the ``on-shell'' contributions are obtained from
 those whose $\delta$ function constraints can be satisfied for
 $\omega \sim k$. It is straightforward to find that \emph{only} the
 terms with $\delta(\omega+p -W_{\vp+\vk})$ have non-vanishing
 support for $\omega \simeq k$. These terms are given in the last
 lines of \ref{PI0} and \ref{PI1} for (sm) and the last lines of \ref{PI0bsm} and \ref{PI1bsm} for
 (bsm) contributions.

 These contributions to the quasiparticle widths in the medium arise
 from the \emph{decay} of the intermediate boson, either the vector
 bosons in the (sm) contributions or the scalars in the (bsm)
 contributions. This is depicted in Fig.(\ref{fig:decay}), the
 Cutkosky cut through the intermediate boson (vector or scalar)
 yields the imaginary part. The process that contributes on shell
 $\omega \simeq k$ is the decay of the boson into the fermions
 (neutrinos and or charged leptons) depicted in this figure.

The fact that the decay of a heavy intermediate state leads to a width was recognized
in ref.\cite{hoboydavey}.

\begin{figure}[h]
\begin{center}
\includegraphics[height=6cm,width=8cm,keepaspectratio=true]{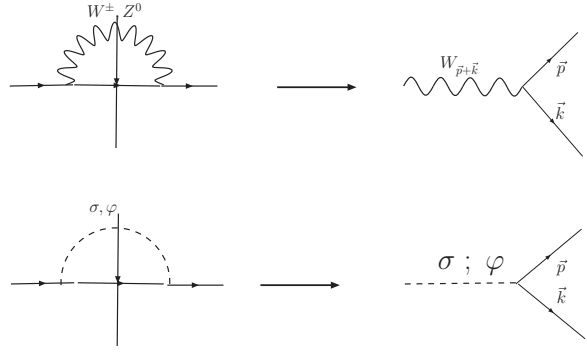}
\caption{ The Cutkosky cut for imaginary part of the (sm) and (bsm)
contributions, and the contribution on the mass shell $\omega \simeq
k$. } \label{fig:decay}
\end{center}
\end{figure}

The analysis of the different cases is simplified by introducing
  \bea \Gamma_{aa} (k_0,k) = && \mathrm{Im} \Bigg[\frac{(k_0+h k)}{2k }\big(\mathds{A}_R
-h\mathds{B}_R \big)_{aa}+\frac{(k_0-h
k)}{2k }\big(\mathds{A}_L+h\mathds{B}_L
\big)_{aa}\Bigg]~,\label{gamaaa}\\
 \Gamma_{ss}(k_0,k) = && \mathrm{Im} \Bigg[\frac{(k_0+h k)}{2k }\big(\mathds{A}_R
-h\mathds{B}_R \big)_{ss}+\frac{(k_0-h
k)}{2k }\big(\mathds{A}_L+h\mathds{B}_L \big)_{ss}\Bigg]\,,
   \label{gamass}\eea in terms of which (see eqn. (\ref{Delta})) \be S_I= 2k \Bigg[\Gamma_{aa}(k_0,k)+\Gamma_{ss}(k_0,k) \Bigg]~~;~~\Delta_I(k_0,k) = \frac{2k}{\delta M^2}
   \Bigg[\Gamma_{aa}(k_0,k)-\Gamma_{ss}(k_0,k) \Bigg] \,.\label{DelIG}\ee

 We need these quantities evaluated on the ``mass shell'', namely for positive energy $k_0 =\overline{\omega}(k) \sim k+\overline{M}^{\,2}/2k$. We find:

 \medskip

 \underline{$\mathbf{h=-1}:$ }  \be  \Gamma_{aa}(k)    \simeq   \mathrm{Im} \big(\mathds{A}_L-\mathds{B}_L
\big)_{aa}~;~ \Gamma_{ss}(k)    \simeq   \mathrm{Im} \big(\mathds{A}_L-\mathds{B}_L
\big)_{ss}~. \label{gammashmin1}\ee

\underline{ $\mathbf{h= 1}:$}   \be  \Gamma_{aa}(k)    \simeq
\mathrm{Im} \Bigg[\big(\mathds{A}_R-\mathds{B}_R \big)_{aa} +
\frac{\overline{M}^{\,2}}{4k^2}~\big(\mathds{A}_L+\mathds{B}_L
\big)_{aa}   \Bigg]~;~ \Gamma_{ss}(k)    \simeq
\mathrm{Im}  \big(\mathds{A}_R-\mathds{B}_R \big)_{ss}~.
\label{gammashplus1}\ee

In the above expressions we have used $Y_{1,2} \ll g$ and $ \overline{M}^{\,2}/4k^2\ll 1$ and
  neglected terms accordingly,   we have suppressed the arguments on $\mathds{A},\mathds{B}$, however,   these matrix elements depend on
   $ k$.
The term with $\mathds{A}_L+\mathds{B}_L$ in (\ref{gammashplus1}) is
noteworthy: the leading contribution to this term is from (sm)
interactions, even setting the Yukawa couplings in the (bsm) sector
to zero a \emph{nearly  right-handed sterile neutrino is produced
via the decay of the vector bosons}.

 The expression for the imaginary parts (\ref{shiftI1},\ref{shiftI2}) simplify in two relevant limits\cite{boymodel}:

 \medskip

 \textbf{a) weak damping: $|\tilde{\gamma}|\ll 1$:} in this limit we
 find \be r\sin(\phi) \simeq \tilde{\gamma} \cos2\theta_m
 \label{WD}\ee leading to the following results for the poles with positive energy \bea \Gamma_1(k) & = &
 \Gamma_{aa}(k)\cos^2\theta_m + \sin^2\theta_m\Gamma_{ss}(k) \label{gama1wd}~,\\\Gamma_2(k) & = &
  \Gamma_{aa}(k)\sin^2\theta_m +
 \cos^2\theta_m\Gamma_{ss}( k) \label{gama2wd}\,.
 \eea Furthermore the difference in the dispersion relations becomes \be  \Delta\Omega_{wd} \equiv \Omega_2(k) - {\Omega}_1(k) \simeq \frac{\delta M^2}{2k} \rho_0  \,,
 \label{freqdiffwd}\ee   which is the usual result for neutrino mixing.

\medskip

\textbf{b) strong damping: ~ $|\tilde{\gamma}| \gg 1$,} in this limit we find \be r\sin(\phi)
 \simeq \tilde{\gamma} ~ \mathrm{sign}(\cos(2\theta_m))~\Bigg[1- \frac{\sin^2(2\theta_m)}{2\tilde{\gamma}^2} \Bigg] \ee
 leading to the following results

 \be  \Gamma_1(k)   =    \frac{1}{2}
\left(\,\Gamma_{aa}(k)+\Gamma_{ss}(k)\,\right)+\frac{1}{2}
\left(\,\Gamma_{aa}(k)-\Gamma_{ss}(k)\,\right)\left(\,\mathrm{sign}(\cos(2\theta_m))-
\frac{\sin^2(2\theta_m)}{2\tilde{\gamma}^2} \,\right)~,
\label{gama1sc}\ee
 \be  \Gamma_2(k)   =    \frac{1}{2}
\left(\,\Gamma_{aa}(k)+\Gamma_{ss}(k)\,\right)-\frac{1}{2}
\left(\,\Gamma_{aa}(k)-\Gamma_{ss}(k)\,\right)\left(\,\mathrm{sign}(\cos(2\theta_m))-
\frac{\sin^2(2\theta_m)}{2\tilde{\gamma}^2}
\,\right)~.\label{gama2sc}\ee In this case the frequency difference between the propagating
states becomes \be  \Delta\Omega_{sd} \equiv  \Omega_2(k) - {\Omega}_1(k) \simeq \frac{\delta M^2}{2k} \rho_0  |\cos(2\theta_m)| =
\frac{\delta M^2}{2k} |\cos(2\theta)+\Delta_R(k)| \,.
 \label{freqdiffsd}\ee This is a remarkable result, the frequency difference \emph{vanishes} at a MSW resonance in striking contrast with the usual quantum mechanics description of neutrino mixing and
 oscillations wherein there is a ``level repulsion'' at an MSW resonance that prevents level crossing.

In all the expressions above $\Gamma_{aa}(k);\Gamma_{ss}(k)$ are given by (\ref{gammashmin1},
\ref{gammashplus1}) in the respective cases $h=\mp1$, and the mixing angle $\theta_m$ is obtained
from eqns.(\ref{mediumangle}) evaluating $\Delta_R$ at $k_0=k$.

The widths for negative energy and $h= \mp 1$ are obtained from the
expressions above by the replacement $\mu \rightarrow -\mu$, this is
a consequence of the relations (\ref{relA},\ref{relB}) and the fact
that the chemical potential is CP-odd, therefore the particle and
antiparticle widths only differ because of the chemical potential.

We emphasize that the results (\ref{gama1wd},\ref{gama2wd}) and
(\ref{gama1sc},\ref{gama2sc}) are \emph{general}, and hold to
\emph{all orders} in perturbation theory as they follow from the
general  form of the self-energies.  In particular these relations
are valid beyond the one-loop order studied here and hold for
\emph{any} processes that contributes to the absorptive parts of the
self-energy at one-loop or higher order.

\subsection{Widths from scalar and vector boson decay:}\label{sec:widths}

As discussed above, the imaginary parts of the self-energy are given in the appendix,
both for (sm) and (bsm) contributions. Inspection of the different
delta functions shows that the \emph{only} contribution
``on-shell'', namely $\omega \simeq k$ arises from   the terms with
$\delta(\omega+p-W_{\vp+\vk})$ in the expressions for the  imaginary
parts (\ref{PI0bsm},\ref{PI1bsm}).

This delta function corresponds to a Cutkosky cut that describes the
process of a scalar   (in (bsm)) or a vector (in (sm))  boson \emph{decay}
into a neutrino and another lepton, displayed in Fig.
(\ref{fig:decay}).

\underline{\textbf{Scalars (bsm):}} For scalars the (R) and (L) components are
the same.  We find for $Y = Y_{1,2};M = M_{\sigma,\varphi}$ for $\sigma,\varphi$ exchange respectively
\be \mathrm{Im}(\mathds{A}_R-\mathds{B}_R) =
\mathrm{Im}(\mathds{A}_L-\mathds{B}_L) = \frac{Y^2 ~ T}{32\pi}
\frac{M^2}{k^2} \ln\Bigg[\frac{1+C_1\,e^{-x^*-\xi}}{1-e^{-x^*-y}}
\Bigg]\,, \label{Imbsmos}\ee
 where \be x^* = \frac{M^2}{4kT}~~;~~\xi=\frac{\mu}{T}~~;~~y =\frac{k}{T} \,,\label{defs2}\ee and
\be C_1 = \Bigg\{ \begin{array}{l}
                  1 ~~\mathrm{for}~~\Sigma^{(1)}_{\sigma,\varphi}~, \\
                  0 ~~\mathrm{for}~~\Sigma^{(2)}_{\sigma,\varphi}~.
                \end{array}  \label{Cdef}\ee

In the relevant region $k < M_{\sigma,\varphi} \sim T$ we can safely neglect the
contribution from the leptonic chemical potential in (\ref{Imbsmos})
and set $\xi =0$, since the bounds from ref.\cite{steigman} suggest
that $|\xi| \lesssim 0.02$.  The result (\ref{Imbsmos}) agrees with that
found in ref.\cite{boysnudm} for the decay of
 the scalar boson into sterile neutrinos (2) ($C_1=0$) for vanishing chemical potential.

 For   $k/T \ll M/T \sim 1$ we can approximate \be \mathrm{Im}(\mathds{A}_R-\mathds{B}_R) =
\mathrm{Im}(\mathds{A}_L-\mathds{B}_L) = \frac{Y^2 ~ T}{32\pi}
\frac{M^2}{k^2}~ e^{-x^*} \Big(C_1+e^{-y}\Big)
 \,. \label{Imbsmosap}\ee

\medskip

\underline{\textbf{Vector bosons (sm):}} For (sm) vector boson exchange (only
L), the imaginary parts ``on-shell'' are obtained from the terms
with $\delta(\omega+p-W_{\vp+\vk})$ in the imaginary parts
(\ref{PI0},\ref{PI1}) setting $\omega \simeq k$. We find \be
\mathrm{Im}(\mathds{A}_L-\mathds{B}_L) = \frac{ g^2_{sm}  T}{16\pi}
\frac{M^2}{k^2} \ln\Bigg[\frac{1+ C_2\,e^{-x^*-\xi}}{1-e^{-x^*-y}}
\Bigg]\,, \label{Imsmos}\ee
 where $g_{sm}$ is given by eqn. (\ref{gsm}),  $M=M_{Z,W}$ for neutral and charged current contributions respectively,
 and \be C_2 = \Bigg\{ \begin{array}{l}
                  1 ~~\mathrm{for}~~\Sigma^{(1)}_{nc},\Sigma_{cc}~, \\
                  0  ~~ \mathrm{for}~~\Sigma^{(2)}_{nc}~.
                \end{array}  \label{Cprimdef}\ee

 For positive energy, and positive helicity (right-handed), we also
need (see eqn. (\ref{gammashplus1}) ) \be
\mathrm{Im}(\mathds{A}_L+\mathds{B}_L) = \frac{ g^2_{sm}
T}{8\pi}\Bigg\{   \ln\Bigg[\frac{1+C_2 \, e^{-x^* }}{1-e^{-x^*-y}}
\Bigg]  + \frac{2T}{k} \Bigg[Li_2\big(e^{-x^*-y}
\big)-C_2\,Li_2\big(-e^{-x^*} \big) \Bigg]\Bigg\}~, \label{ImApBsm}\ee
where $Li_2$ is the dilogarithm or Spence's function and we have set
$\mu=0$. This expression simplifies in the limit $k/T \ll M/T \sim 1$ with the result \be \mathrm{Im}(\mathds{A}_L+\mathds{B}_L) \simeq\frac{ g^2_{sm}
T^2}{4\pi k}~ e^{-x^*} \Big(C_2+e^{-y}\Big)~. \label{Implusap}\ee In the above results for vector
bosons $M=M_{W,Z}$ respectively.

We can now gather all the results needed for $\Gamma_{aa}(k);\Gamma_{ss}(k)$ (\ref{gammashmin1},\ref{gammashplus1})
and the quasiparticle widths $\Gamma_{1,2}(k)$ obtained from them. Approximating $\cos(\theta) \sim 1;
\sin(\theta)\sim 0$ we find,

\be \mathrm{Im}\big(\mathds{A}_R-\mathds{B}_R \big)_{aa} = \frac{Y^2_1 T}{32\pi}~
\Big( \frac{M^2_\sigma }{k^2}\Big)~\ln\Bigg[ \frac{1}{1-e^{-x^*_\sigma}~e^{-y}}\Bigg]\,; \label{ImRaa}  \ee

 \be  \mathrm{Im}\big(\mathds{A}_R-\mathds{B}_R \big)_{ss} = \mathrm{Im}\big(\mathds{A}_L-\mathds{B}_L \big)_{ss}   =
 \frac{Y^2_1 T}{32\pi}~
\Big( \frac{M^2_\sigma }{k^2}\Big)~\ln\Bigg[ \frac{1+e^{-x^*_\sigma}}{1-e^{-x^*_\sigma}~e^{-y}}\Bigg] +
 \frac{Y^2_2 T}{32\pi}~
\Big( \frac{M^2_\varphi }{k^2}\Big)~\ln\Bigg[ \frac{1 }{1-e^{-x^*_\varphi}~e^{-y}}\Bigg];  \label{ImRLss}     \ee

\bea    \mathrm{Im}\big(\mathds{A}_L-\mathds{B}_L \big)_{aa}   & =
& \frac{g^2 T}{32\pi}~\Bigg\{
\frac{1}{2 \cos^2(\theta_w)}\Big( \frac{M^2_Z }{k^2}\Big)\ln\Bigg[ \frac{1+e^{-x^*_Z}}{1-e^{-x^*_Z}~e^{-y}}\Bigg] +
 \Big( \frac{M^2_W }{k^2}\Big)~ \ln\Bigg[ \frac{1+e^{-x^*_W} }{1-e^{-x^*_W}~e^{-y}}\Bigg]\Bigg\} \nonumber \\& & +\frac{Y^2_1 T}{32\pi}~
\Big( \frac{M^2_\sigma }{k^2}\Big)~\ln\Bigg[ \frac{1}{1-e^{-x^*_\sigma}~e^{-y}}\Bigg]; \label{ImLaa}  \eea

\bea    \mathrm{Im}\big(\mathds{A}_L+\mathds{B}_L \big)_{aa}  & = &
 \frac{g^2 T}{16\pi} ~\Bigg\{\frac{1}{2\cos^2(\theta_w)}\Bigg[\ln\Big( \frac{1+e^{-x^*_Z}}{1-e^{-x^*_Z}~e^{-y}}\Big) + \frac{2T}{k}
 \Big[Li_2\big( e^{-x^*_Z}~e^{-y}\big)-Li_2\big(-e^{-x^*_Z} \big)\Big] \Bigg]\nonumber \\ & & +
 \ln\Big( \frac{1+e^{-x^*_W} }{1-e^{-x^*_W}~e^{-y}}\Big)+ \frac{2T}{k} \Big[Li_2\big(  e^{-x^*_W}~e^{-y}\big)-Li_2\big(-e^{-x^*_W} \big)\Big]\Bigg\} +\frac{Y^2_1 T}{16\pi}\Big(\frac{2T}{k}\Big)Li_2\big(  e^{-x^*_{\sigma}}~e^{-y}\big). \nonumber \\
  \label{ImPLaa}  \eea

 In the expressions above we have defined \be x^*_\alpha = \frac{M^2_\alpha}{4kT}~~;~~\alpha=\sigma,\varphi,Z,W \,.\label{xs}\ee For small values of the arguments $Li_2(z) \sim z$ which may be used appropriately whenever $x^*_\alpha >1  $, a situation which describes
 the relevant range $M_\alpha \sim T;k<T$.

 Eqns (\ref{ImRaa}-\ref{ImPLaa}) combined with (\ref{gammashmin1},\ref{gammashplus1}) yield the complete expressions for
 the quasiparticle widths $\Gamma_{1,2}$ in all cases, and as per the discussion below, the production rates.

\subsection{Imaginary parts: from the width to the production rates.}

The connection between the quasiparticle widths (imaginary part of the self-energy ``on-shell'')
and the production rate is established via the Boltzmann equation for the production of a given
species, in this case that of a ``sterile'' neutrino. Consider as an example the scalar vertex $Y_1 \overline{\nu}_s\,\sigma \nu_a$,
 the analysis is similar for the other, including (sm) vertices. The Boltzmann
equation is of the form (gain)$-$ (loss) (see for example the appendix in ref.\cite{boysnudm})). The
gain term corresponds to the \emph{decay} process $\sigma\rightarrow \overline{\nu}_a+ \nu_s$ and is of the
form\cite{boysnudm} \be \frac{dn_s(k)}{dt}\Bigg|_{gain} =  \int \frac{d^3p}{(2\pi)^3} \Big|\mathcal{M}_{fi}\Big|^2 ~\delta \big(W_{\vp+\vk} - p - k\big) N_B\big(W_{\vp+\vk}\big)(1-\overline{n}_F(p))(1-n_s(k))~, \label{gain} \ee where $N_B, n_F$ are the bosonic and fermionic distribution functions respectively. The loss term describes the inverse process, namely the recombination $\overline{\nu}_a + \nu_s \rightarrow \varphi$ with
\be \frac{dn_s(k)}{dt}\Bigg|_{loss} = \int \frac{d^3p}{(2\pi)^3} \Big|\mathcal{M}_{fi}\Big|^2~ \delta\big(W_{\vp+\vk} - p - k\big)  \Big[1+N_B\big(W_{\vp+\vk}\big)\Big]\overline{n}_F(p) n_s(k) \,. \label{loss} \ee
Therefore the Boltzmann equation is of the form
\be \frac{dn_s(k)}{dt}  = \int \frac{d^3p}{(2\pi)^3} \Big|\mathcal{M}_{fi}\Big|^2~
 \delta\big(W_{\vp+\vk} - p - k\big) \Bigg\{ N_B\big(W_{\vp+\vk}\big)(1-\overline{n}_F(p))(1-n_s(k))- \Big[1+N_B\big(W_{\vp+\vk}\big)\Big]
 \overline{n}_F(p) n_s(k)\Bigg\}~. \label{boltz}\ee
If the distribution function of the particle in question is slightly perturbed \emph{off
equilibrium}, the relaxation rate of the distribution function towards equilibrium is obtained by writing $n_s(k) = n^{eq}_s(k) +\delta n_s(k)$
 and linearizing the Boltzmann equation in $\delta n_s(k)$\cite{nosfermions}.
The linearized Boltzmann equation reads \be \frac{d\delta n_s(k)}{dt} = -\Gamma_{rel} ~\delta n_s(k)~, \label{linboltz}\ee
where \be \Gamma_{rel} = \int \frac{d^3p}{(2\pi)^3} \Big|\mathcal{M}_{fi}\Big|^2~ \delta\big(W_{\vp+\vk} - p - k\big)
 \Big[\overline{n}_F(p)+N_B\big(W_{\vp+\vk}\big)\Big]~. \label{gamrel}\ee
 As discussed in ref.\cite{nosfermions}, the relaxation rate $\Gamma_{rel}$ is \emph{twice}
 the quasiparticle width\cite{nosfermions} since the distribution function
is bilinear in the fields. The relation between $\Gamma_{rel}$  and
the on-shell width becomes evident comparing the expression
(\ref{gamrel}) with the ``on-shell'' imaginary parts, namely the
last lines in eqns. (\ref{PI0},\ref{PI1},\ref{PI0bsm},\ref{PI1bsm})
with $\omega \simeq k$. The production rate of the sterile species
is obtained by neglecting the inverse process and neglecting the
sterile population buildup in the Boltzmann equation (\ref{boltz}),
namely, \be \frac{dn_s(k)}{dt}\Bigg|_{prod}  = \int
\frac{d^3p}{(2\pi)^3} \Big|\mathcal{M}_{fi}\Big|^2~
 \delta\big(W_{\vp+\vk} - p - k\big)  N_B\big(W_{\vp+\vk}\big)(1-\overline{n}_F(p))~.
 \label{prods}\ee Therefore by obtaining the bosonic and fermionic
 contributions to the quasiparticle widths as in the previous
 section, we can obtain the production rate. Although the term with
 the product $N_B \overline{n}_F $ is not included in the width,
 such term is smaller than the term with $N_B$ only, since $p^2 \overline{n}_F(p)$ features a maximum at
 $p/T\sim 2.3$ for which $\overline{n}_F(p) \sim 0.09$ (for $\mu/T \ll 1$). Therefore  in the region of importance in
 the integral $p\gtrsim T$,   the production and relaxation rates
 only differ by a few percent, and the results for the relaxation
 rates yield a  reliable approximation to the production rate.

 An important bonus of obtaining the production rate from the
 quasiparticle decay width as carried out here  is the correct dependence on the mixing
 angle in the medium, which would be missed by a naive perturbative
 calculation.

 Therefore the quasiparticle width yields an
 excellent approximation to the production rate, in particular it
 describes correctly the dependence on the mixing angles in the
 medium,   its  magnitude and $k$-dependence.

 In particular, the result (\ref{ImRLss}) confirms the result of
 ref.\cite{boysnudm} for $Y_1=0$. For the scalar contribution  (bsm) the right and
 left-handed yield the same result (multiplying (\ref{ImRLss}) by a
 factor 2 in the total rate) and as discussed above the production rate is twice the
 width, which restores the factor 4 between (\ref{ImRLss}) and the
 result in ref.\cite{boysnudm} which corresponds to the case
 $Y_1=0$.

 Thus we conclude that the results of eqns. (\ref{gammashmin1},\ref{gammashplus1},\ref{gama1wd},\ref{gama2wd},\ref{gama1sc},\ref{gama2sc})
 along with the explicit forms (\ref{ImRaa}-\ref{ImPLaa}) provide a
 complete and reliable assessment of the production rates ready to
 be input in the kinetic equations that include the cosmological
 expansion\cite{boysnudm}.

 \subsection{Weak or strong damping?} We have now all the ingredients to assess under which
 circumstances the weak ($|\tilde{\gamma} | \ll 1$)  or strong ($|\tilde{\gamma}| \gg 1$) damping conditions are fulfilled. In terms of the widths and real parts it follows that
\be \tilde{\gamma}
     \simeq \frac{2k}{  M^2_s}\frac{\Big[\Gamma_{aa}(k)-\Gamma_{ss}(k) \Big]}{ \Bigg[\Big(\cos(2\theta)+  {\Delta_R(k)}  \Big)^2
+\sin^2(2\theta)\Bigg]^\frac{1}{2}}~.  \label{fintilga}\ee

For $h=-1$,  $\Gamma_{aa}-\Gamma_{ss}$ and $\Delta_R$ are dominated by the (sm) contributions, therefore
from eqns.(\ref{DelRhm1ap}) and (\ref{ImLaa}) we find
\bea \Delta_R(k)  & \sim &  \frac{g^2}{16\pi^2} \frac{kT}{M^2_s}~ \mathcal{A}(k)~,\label{delRap}\\
\Delta_I(k) & \sim & \frac{g^2}{32\pi } \frac{kT}{M^2_s}~\Bigg(
\frac{M_Z}{k}\Bigg)^2 ~ \mathcal{B}(k)\,, \label{delIap}\eea
where
$\mathcal{A}(k),\mathcal{B}(k)$ can be read off
(\ref{DelRhm1ap},\ref{ImLaa}). In the region of parameters where
$\Delta_R(k) \gg \cos(2\theta) \sim 1$, it follows that
$\tilde{\gamma} \simeq \Delta_I(k)/\Delta_R(k)$, furthermore,  for
$k < T\sim M_{Z,W}$ the function $\mathcal{B}(k)\sim e^{-x^*_Z} \ll
1$, leading to  $\Delta_I/\Delta_R \ll 1$ corresponding to the weak
damping case in which the widths (production rates) are given by
(\ref{gama1wd},\ref{gama2wd}).

Far away from the MSW resonances but in the region where
$\cos(2\theta) \sim 1 \gg \Delta_R(k)$ it also follows that
 $ \Delta_I/\Delta_R \ll 1$, corresponding again to
 the weak damping regime. Therefore the parameter region \emph{far away} from MSW resonances (either above or below)
 corresponds to the weak  damping regime.

 Very near MSW resonances $\cos(2\theta)+\Delta_R \sim 0$ and $\tilde{\gamma} \sim \Delta_I/|\sin(2\theta)|$,
 in the region of relevance for our analysis $T\sim M_{Z,W}$ with $M_s \sim \mathrm{KeV}$
 it follows that \be \frac{\Delta_I(k)}{|\sin(2\theta)|} \sim \frac{4\times 10^{13}}{|\sin(2\theta)|}~\Big( \frac{k}{T} \Big) ~\Bigg( \frac{M_Z}{k}\Bigg)^2
  ~ \mathcal{B}(k)~, \label{biggi}\ee therefore, since the resonance occurs at $k/T <  1 $ for $M_{Z,W}\sim
  T$ we conclude that
  the strong damping condition $\tilde{\gamma} \gg 1$ is fulfilled
 \emph{near MSW resonances}. Because the MSW resonance(s) are very narrow
 for $T\simeq M_{Z,W}$ as discussed above (see the discussion leading to eqn. \ref{sharpmsw}), we conclude that in \emph{most} of the regime of temperatures and momenta
 the weak damping results (\ref{gama1wd},\ref{gama2wd}) are valid and only in a very narrow region near
 MSW resonances the strong damping results (\ref{gama1sc},\ref{gama2sc}) are valid.

An identical analysis confirms a similar conclusion for the case
$h=1$, namely the weak damping condition holds in most of the
relevant range of $M/T;k/T$ but for a narrow region near the MSW
resonances in which the strong damping condition holds.

An alternative interpretation of the weak and strong damping regime
is obtained  using eqn. (\ref{freqdiffwd}) to write  \be
\tilde{\gamma} \simeq \frac{\Gamma_{aa}-\Gamma_{ss}}{\Delta
\Omega_{wd}}~.  \ee Since $\Delta \Omega_{sd} \leq \Delta \Omega_{wd}$
the denominator gives an upper bound to the \emph{oscillation
frequency} between the active and sterile neutrinos. The weak
damping regime $|\tilde{\gamma}| \ll 1$ describes the case in which
there are many oscillations before the overlap amplitude is
suppressed, whereas the strong damping regime describes the case in
which damping occurs before oscillations take place. For a similar
discussion see the second ref. in\cite{hoboya}.

\subsection{Regime of validity of perturbation theory.}\label{subsec:PT}

In the relativistic approximation the validity of the perturbative expansion requires that  $k \gg \Sigma_{bsm};\Sigma_{sm}$. Since the weak interaction coupling constant $g_{sm}$ is much larger than $Y_{1,2}$ we focus on the standard model contributions.

From the expression (\ref{ReSigsm}) and the results displayed in fig. (\ref{fig:coeffuncssm}) we see that for $M_{W}/T \gtrsim 1$, it follows that $ \mathrm{Re}\Sigma_{sm}\propto \alpha_w T$ since the coefficient functions $A,B \lesssim 12$. Therefore perturbation theory is valid for $k \gg T/30$, hence for $M_W/T \gtrsim 1$ the resonances in absence of lepton asymmetry at $0.2 \lesssim k/T \lesssim 1$ for $1 \lesssim M_W/T \lesssim 3$    are comfortably within the regime of vali
dity of the perturbative expansion.
The lepton-asymmetry induced resonance for $k/T  \ll M_W/T$ is
the usual resonance and for $T\ll M_{W}$ the expressions (\ref{sig1nc}-\ref{sigcc})
 reduce to the results available in the literature\cite{notzold,boyhonueu}.
 In the regime $k\ll T < M_W$ the on-shell self-energies  are linear in $k$.
 We see that, for $g^2 \sim 0.4$,  the terms proportional to $k$ are $\ll 1$ for $M_{W} \gtrsim 2T$,
 hence perturbation theory is reliable    within the regime of interest in this article.
  The imaginary parts are always perturbatively small because of the exponential suppression factors $e^{-M^2/kT}$.

  Perturbation theory breaks down for  $M\lesssim T$ for the small $k/T$ region and requires a hard thermal loop resummation
  program\cite{htl} akin to the one presented in ref.\cite{boyaneutrino} in the standard model without mixing.  This is well known in gauge theories where the gauge bosons are nearly massless on the scale $T$\cite{htl}. Such program   is  well beyond the realm of this study, however for $M/T \gtrsim 1-3$ our results are reliable for $k/T \gg \alpha_w$ as analyzed above. For example for the case $M/T \sim 1$ although the peak in the coefficients $A, B$ in the self-energy occur for $k/T \approx 0.07$ which is not too large compared to $\alpha_w \sim 0.03$, the position of the resonance at $k/T \sim 0.2$ is well within the regime of validity of the perturbative expansion. The validity of perturbation theory improves
  dramatically for $M/T > 1$ even in the low momentum region as discussed above. Therefore, we conclude that for $M/T >1$ the perturbative results are reliable for
  $k/T > \alpha_w$, in particular the new resonances are well within the regime of
  validity of the perturbative expansion. The results for the production rates are always
  perturbatively small and reliable because of the exponential suppression factor.

 \section{Discussion}\label{sec:discussion}
 Our goal is to study the production of sterile neutrinos in cosmology near
 the electroweak scale when the universe is radiation dominated. To include the effects
 of cosmological expansion in the production rates and mixing angles, one must first replace
 the momentum $k\rightarrow k_p(t) =k /a(t)$ and temperature $ T \rightarrow T(t) = T_i a_i/a(t)$ where $k$ is the  comoving momentum,
 $a(t)$ the scale factor and $T_i;a_i$ correspond to the initial temperature  and scale factor at which the kinetic equations are initialized.
 Whereas the ratio $k_p(t)/T(t) = k/(T_i a_i)$ is constant  $M/T(t) = M a(t)/(T_i a_i)$ \emph{grows} during
 the expansion. Consider setting initial conditions at $T_i \lesssim M_W$, so that $M/T_i \sim 1$,
 the analysis of section (\ref{sec:realparts}) shows that there exists at least one very narrow MSW resonance even for
 \emph{nearly right-handed sterile neutrinos} (two if
 a lepton asymmetry in the neutrino sector is included) at a value $\big(k_p(t)/T(t)\big)_c < 1$. For
$\big(k_p(t)/T(t)\big)< \big(k_p(t)/T(t)\big)_c $ the analysis shows that $\Delta_R \gg 1$ and
\be \theta_m \sim \frac{\theta}{ \Delta_R } \ll \theta \,, \label{tetas}\ee therefore for $\big(k_p(t)/T(t)\big)< \big(k_p(t)/T(t)\big)_c $ we find
\bea \Gamma_1 & \sim & \Gamma_{aa}~,\nonumber \\ \Gamma_2 & \sim &  \Gamma_{ss} + \Big(\frac{\theta}{ \Delta_R } \Big)^2\Gamma_{aa} \,. \label{ratess}\eea   For these values of $k_p(t)/T(t)$ the mode ``1'' is active-like and it is produced with a weak interaction rate, whereas the mode ``2'' is
sterile like and is produced with the rate similar to that of ref.\cite{boysnudm} plus small
corrections  from standard model interaction rates suppressed by the mixing angle in the medium $\sim \theta/\Delta_R$ . On the other hand for $\big(k_p(t)/T(t)\big)> \big(k_p(t)/T(t)\big)_c $ we found above that $\Delta_R \ll -1$ leading to $\theta_m \sim \pi/2$, namely the mode ``1'' is sterile-like and the mode ``2'' is active like, with the production rates
\bea \Gamma_1 & \sim &\Big(\frac{\theta}{2 \Delta_R} \Big)^2 \Gamma_{aa}+\Gamma_{ss}~,\nonumber \\ \Gamma_2 & \sim &  \Gamma_{aa}   \,. \label{ratesl}\eea As the cosmological expansion proceeds eventually $M/T(t) \gg 1$
and the resonances disappear (in absence of lepton asymmetry the MSW resonances for $k_p(t)/T(t) < 1$
disappear for $M/T(t) \gtrsim 3$), $\Delta_R$ remains large but positive and the mixing angle in the
medium is given by (\ref{tetas}) and the production rates are given by (\ref{ratess}) for all values
of $k_p(t)/T(t)$, namely the mode ``1'' remains the active-like and the mode ``2'' the sterile-like.

We note that (see section (\ref{sec:widths}))
\be \Gamma_{ss},\Gamma_{aa} \propto \Big(\frac{M^2}{k^2}\Big) \ln\Big[\frac{1}{1-e^{-x^*}e^{-y}} \Big] \,, \label{prefack}\ee this is precisely the form
of the production rate that leads to a distribution function after freeze-out that is enhanced at
small momentum, a feature that leads to a larger free streaming length and transfer
function at small scales\cite{boysnudm}.

During the time when $M/T(t) \sim 1$ the MSW resonance for $k_p(t)/T(t) <1$ leads to a \emph{non-thermal}
population of neutrinos: for $\big(k_p(t)/T(t)\big)< \big(k_p(t)/T(t)\big)_c $ there is a large
production of mode ``1'' leading to large populations and a small production of ``2'' (sterile like)
leading to small populations, whereas for $\big(k_p(t)/T(t)\big) > \big(k_p(t)/T(t)\big)_c $ there is a
``population inversion'' in the sense that mode ``1'' is slightly populated whereas mode ``2'' will be
substantially populated, however, without the small momentum enhancement. Consider  a fixed  value of $k_p(t)/T(t) <1$ during the cosmological expansion the ratio $M/T(t) \propto a(t)$ increases sweeping through the MSW resonance, when this happens the mixing angle in the medium vanishes very rapidly because
the resonance is very narrow and the mode ``2'' becomes sterile like.   As the expansion continues the MSW resonances (in absence of lepton
asymmetry) disappear altogether  and the mixing angles and production rates are given by (\ref{tetas},\ref{ratess})
respectively for \emph{all values} of $k_p(t)/T(t)$. The population of the active-like
neutrino (mode ``1'') continues to build up via weak interaction processes, including those that become
 dominant at $T \ll M_{W}$ and eventually thermalizes,
whereas the population of the sterile-like neutrino will be frozen-out as the production rate $\Gamma_2$
shuts-off as $\Gamma_{ss} $ vanishes rapidly for $M/T(t) \ll 1$ (see ref.\cite{boysnudm}) and
$\theta_m \rightarrow 0$ as $M/T(t) \gg 1$ even when $\Gamma_{aa}$ (weak interaction rates) remain
large down to the decoupling temperature of weak interactions $\sim 1 ~ \mathrm{MeV}$.

This analysis indicates that sterile neutrino production via the decay of scalar or vector bosons will be effective only in a   region
for $M_W/T(t) \sim 1$ and the distribution function at freeze-out will be \emph{strongly non-thermal} with
very small population but with
an enhancement at small momentum  as found in ref.\cite{boysnudm}. However, the weak interaction contribution will  freeze out much later,
depending on the temperature dependence of the mixing angle in the medium and will eventually merge
with the non-resonant (DW) production mechanism\cite{dw} at $T \sim 150 \mathrm{MeV}$.

However the non-thermal distribution built up during the stage when scalar and vector boson decay dominate the production will remain.

 At this stage it is important to understand the self-consistency of the analysis. In obtaining the
 self-energies we had assumed that the eigenstate ``1'' is active like with a thermal distribution
 function. We have learned, however, that there are resonances and the eigenstates ``1'' and ``2''
 are either active-like or sterile-like depending on $k$, namely on which side of the MSW resonance
 the wavevector lies. This finding calls into question the thermal nature of the neutrino propagator
 in the intermediate state (of course there is no such ambiguity in the charged lepton propagator that
 enters in the charged current self-energy). This issue notwithstanding, we have found that the
fermionic and bosonic contributions to the real parts of the
self-energies are \emph{qualitatively the same} with a rather small
quantitative difference, both for (sm) and (bsm) contributions.
Therefore replacing the thermal fermion propagator for a vacuum one
leads to a minor quantitative modification of our arguments. However
because of the enormous pre-factors the conclusions about the
sharpness of the resonance and the resonance positions  \emph{do not
change} and the general analysis remains the same. Therefore, we
conclude that the results obtained above are very \emph{robust} not
depending on whether the intermediate fermion line features a
thermal or vacuum propagator or non-thermal propagator interpolating
between these two cases.

 \section{Conclusions and cosmological consequences}

 A comprehensive program to assess the viability of any potential (DM) candidate begins
 with the microphysics of the production and freeze-out process of the particle physics
 candidate. This initial step determines the distribution function at freeze-out which
 in turn determines, along with the mass, its abundance, free streaming
 length, phase space density at decoupling and the transfer function and power spectrum in the linear regime. Our
 objective is to carry out this program for sterile neutrinos with mass in the $\mathrm{KeV}$
 range which seems to be the range favored not only as a (DM) candidate but also provide potential
 solutions to  a host of astrophysical problems\cite{kusenko}.

 In this article we focus on the first step of the program and study  the production of sterile neutrinos in a temperature regime  near the electroweak scale  in an extension beyond the standard model in which the see-saw
 mass matrix emerges from expectation values of Higgs-like scalars with masses of the order
 of the electroweak scale. This simple and compelling extension  which features only one scale yields
  rich phenomenology\cite{kusenko,shapo,kusepetra}. The main observation in this article
 is that in this temperature range sterile neutrinos are produced by the decay not only of the
 Higgs-like scalar as explored in refs.\cite{kusepetra,boysnudm} but also of the \emph{charged and
 neutral vector bosons} of the standard model. We consider active and sterile species to be Dirac
 fermions to allow the possibility to include a lepton asymmetry hidden in the (active) neutrino
 sector consistent with recent bounds from WMAP and BBN\cite{steigman}.

 The assessment of the contribution from standard model vertices to sterile neutrino production requires
 an analysis of the mixing angles in the medium and production rates. We obtain both from the study
 of the full equation of motion of the active and sterile neutrinos that input the self-energies in the
 medium. The real part of the self-energy (index of refraction) determines the dispersion relations and
 mixing angles in the medium, and the imaginary (absorptive) part determines the production rates.

  We provide a detailed analysis of the contributions from ``beyond the standard model'' and standard
 model interactions to the mixing angles, dispersion relations and production rates, thereby facilitating
 the analysis of different situations. The study of the ``index of refraction'' in the temperature
 regime near the electroweak scale has not been performed before and yields a wealth of remarkable
 phenomena.

 Our study reveals the presence of narrow MSW resonances \emph{even in the absence of a lepton asymmetry},
 in the temperature regime $ T \gtrsim M_{W}$ for $k/T \lesssim 1$.   For vanishing lepton asymmetry
 the resonance occurs at a value $(k/T)_c$ that depends on the ratio
 $M_W/T$ with $0.15 \lesssim (k/T)_c \lesssim 1$ for $1 \lesssim (M_W/T) \lesssim 3$.
 The position of the resonance  $(k/T)_c$ increases with $M_W/T$,
 the resonance eventually disappear for $M_W \gg T$ recovering the result valid in the
 Fermi limit of the weak interactions\cite{notzold,boyhonueu}.

 Including the possibility of
 a (small) lepton asymmetry in the neutrino sector with a value compatible with the bounds from
 WMAP and BBN\cite{steigman} yields \emph{two} narrow MSW resonances in these regions, with the resonance
 associated with the lepton asymmetry occurring at $k <  \mu \ll T$ where $\mu$ is the chemical
 potential for the active species that determines the lepton asymmetry.

  A remarkable
 aspect of these results is that near these resonances the contribution of the imaginary part of
 the self-energies leads to a strong damping regime, and the difference in the propagating frequencies
 \emph{vanishes} exactly at the position of the resonance, with a concomitant breakdown of adiabaticity.
 For $M_{W} \gg T$   the MSW resonances that are independent of the lepton asymmetry  disappear
 leaving only the low energy resonances associated with  the lepton asymmetry.

 Furthermore, we have found that it is quite possible that the region of parameters of the extension
 (bsm) allow for MSW resonance for positive energy, positive helicity, namely nearly \emph{right-handed}
 states both with and without lepton asymmetry. We also find that
 the decay of the $Z^0,W^{\pm}$ vector bosons leads to the
 production of nearly \emph{right-handed} sterile-like neutrinos.

  Because the resonances are very narrow, we obtain a simple expression for the production rates
   (see section \ref{sec:discussion}) that  is valid in a wide range of temperatures and
 clearly displays the contribution from standard model and beyond standard model interactions.

 We have argued that in the early universe the cosmological expansion leads to a highly \emph{non-thermal}
 distribution function for sterile neutrinos with an enhancement of the low momentum region $k< T$ both as a consequence of the
  MSW resonances and the vanishing of the mixing angle and production rates as the temperature falls well below the electroweak
   scale.  Furthermore, we expect that because the
 MSW resonances are very narrow, the cosmological expansion will lead to   sterile neutrino
 production resulting in a highly non-thermal distribution with low momentum enhancement. The form
 of the production rates via scalar and vector boson decay are similar to that in ref.\cite{boysnudm},
 which leads us to conjecture that the distribution function after freeze-out will be enhanced in the low
 momentum region, leading to a smaller free streaming length and larger power spectrum at small scales as compared
 to the (DW) mechanism\cite{dw,boysnudm}.

 The next step of the program will input these results into the kinetic equations that describe the
 production and freeze-out of the sterile species from which the distribution function at decoupling
 is obtained. We expect to report on these studies in a forthcoming article.

An important remaining question is the extrapolation of these
results to $T \gg  M_{W,Z}$.   At temperatures above the
 electroweak symmetry breaking scale the $SU(2)\times U(1)$
symmetry is restored and the vector bosons  become massless at tree
level, therefore the production
 channel described here shuts off. However, vector bosons
 acquire electric screening masses of order $gT$\cite{htl} and scalar bosons may also acquire thermal mass
 corrections of $\mathcal{O}(Y_{1,2} T)$.

Furthermore the see-saw mass matrix also vanishes at tree level if
all the mass terms arise from the expectation value of the
Higgs-like scalar field.  This high
 temperature regime requires a deeper understanding of radiative corrections to the propagators of the
 vector bosons, in particular the hard-thermal loop
 corrections\cite{htl}.

  Understanding the possibility of sterile
 neutrino production in this high temperature regime entails a non-perturbative resummation program
 also for neutrinos, akin to the study in ref.\cite{boyaneutrino}.
 This program although clearly interesting in its own  right is far beyond the realm of our goals here and deserves a deeper study.

\appendix
\section{Vector boson exchange (SM)} The (SM) self-energy contributions with the exchange of a vector boson are given by the
spectral representation (\ref{selfa}) with the imaginary part given
by eqn. (\ref{imsigsm}) which is of the form \be
Im\Sigma_{sm}(\omega,\vk)=       \frac{ \pi g^2_{sm} }{4}   \int
\frac{d^3p}{(2\pi)^3\,p\,W_{\vp+\vk} }\Big[\gamma^0
\,\Pi^0_{sm}(\omega,\vp,\vk)- \vec{\gamma}\cdot \hat{k}\,
\Pi^1_{sm}(\omega,\vp,\vk)\Big] \,. \label{Imsigdec}\ee Neglecting
the mass of the neutrinos and charged leptons we find

\bea \Pi^0_{sm}(\omega,\vp,\vk)    & = &
\Big[1-n_F(p)+N_B(W_{\vp+\vk})\Big] \Big[p \big(1+ \frac{2
W^2_{\vp+\vk}}{M^2} \big) +
 \frac{2W_{\vp+\vk}}{M^2}\big(p^2 + \vk\cdot\vp \big) \Big]~\delta\big(\omega-p-W_{\vp+\vk}\big) \nonumber \\
  &  +  &
\Big[1-\overline{n}_F(p)+N_B(W_{\vp+\vk})\Big] \Big[p \Big(1+
\frac{2 W^2_{\vp+\vk}}{M^2} \Big) + \frac{2W_{\vp+\vk}}{M^2}\Big(p^2
+ \vk\cdot\vp \Big) \Big]~\delta\Big(\omega+p+W_{\vp+\vk}\Big)
\nonumber \\ &  + & \Big[ n_F(p)+N_B(W_{\vp+\vk})\Big] \Big[p
\Big(1+ \frac{2 W^2_{\vp+\vk}}{M^2} \Big) -
\frac{2W_{\vp+\vk}}{M^2}\Big(p^2 + \vk\cdot\vp \Big)
\Big]~\delta\Big(\omega-p+W_{\vp+\vk}\Big) \nonumber \\ &  +  &\Big[
\overline{n}_F(p)+N_B(W_{\vp+\vk})\Big] \Big[p \Big(1+ \frac{2
W^2_{\vp+\vk}}{M^2} \Big) - \frac{2W_{\vp+\vk}}{M^2}\Big(p^2 +
\vk\cdot\vp \Big) \Big]~\delta\Big(\omega+p-W_{\vp+\vk}\Big)
\label{PI0}\eea  and

\bea \Pi^1_{sm}(\omega,\vp,\vk)    & = & \Big[1-n_F(p)+N_B(W_{\vp+\vk})\Big] \Big[ -\hat{k}\cdot \vp+
\frac{2\Big( k+\hat{k}\cdot\vp \Big)}{M^2} \Big(p\,W_{\vp+\vk}+p^2+\vk\cdot\vp \Big)\Big]~\delta\big(\omega-p-W_{\vp+\vk}\big) \nonumber \\
  &  -  &
\Big[1-\overline{n}_F(p)+N_B(W_{\vp+\vk})\Big] \Big[ -\hat{k}\cdot
\vp+ \frac{2\Big( k+\hat{k}\cdot\vp \Big)}{M^2}
\Big(p\,W_{\vp+\vk}+p^2+\vk\cdot\vp
\Big)\Big]~\delta\big(\omega+p+W_{\vp+\vk}\big) \nonumber \\ &  + &
\Big[ n_F(p)+N_B(W_{\vp+\vk})\Big] \Big[ -\hat{k}\cdot \vp+
\frac{2\Big( k+\hat{k}\cdot\vp \Big)}{M^2}
\Big(-p\,W_{\vp+\vk}+p^2+\vk\cdot\vp
\Big)\Big]~\delta\big(\omega-p+W_{\vp+\vk}\big) \nonumber \\ &  -  &
\Big[ \overline{n}_F(p)+N_B(W_{\vp+\vk})\Big] \Big[ -\hat{k}\cdot
\vp+ \frac{2\Big( k+\hat{k}\cdot\vp \Big)}{M^2}
\Big(-p\,W_{\vp+\vk}+p^2+\vk\cdot\vp
\Big)\Big]~\delta\big(\omega+p-W_{\vp+\vk}\big)~. \label{PI1}\eea

\section{Scalar exchange (BSM)}

 For scalar boson exchange we find  \be \mathrm{Im}\Sigma_{bsm}(\omega,\vk) =
 \frac{\pi  Y^2}{4}\int \frac{d^3p}{(2\pi)^3\,W_{\vp+\vk}}\Big[\gamma^0 \,
 \Pi^0_{bsm}(\omega, \vp,\vk)-\vec{\gamma}\cdot \hat{k} ~\Big(\hat{k}\cdot \hat{p} \Big)~ \Pi^1_{bsm}(\omega,\vp,\vk)\Big]~, \label{imsigbsm}\ee where

 \bea \Pi^0_{bsm}(\omega, \vp,\vk) &  = &  \Big[1-n_F(p)+N_B(W_{\vp+\vk})\Big]~
 \delta\big(\omega-p-W_{\vp+\vk}\big) + \Big[1-\overline{n}_F(p)+N_B(W_{\vp+\vk})\Big]~\delta\Big(\omega+p+W_{\vp+\vk}\Big) \nonumber \\
 & + & \Big[ n_F(p)+N_B(W_{\vp+\vk})\Big]~\delta\big(\omega-p+W_{\vp+\vk}\big) +
 \Big[ \overline{n}_F(p)+N_B(W_{\vp+\vk})\Big] ~\delta\big(\omega+p-W_{\vp+\vk}\big)~, \label{PI0bsm}\eea

  \bea \Pi^1_{bsm}(\omega, \vp,\vk) &  = &  \Big[1-n_F(p)+N_B(W_{\vp+\vk})\Big]~ \delta\big(\omega-p-W_{\vp+\vk}\big) -
  \Big[1-\overline{n}_F(p)+N_B(W_{\vp+\vk})\Big]~\delta\Big(\omega+p+W_{\vp+\vk}\Big) \nonumber \\
 & + & \Big[ n_F(p)+N_B(W_{\vp+\vk})\Big]~\delta\big(\omega-p+W_{\vp+\vk}\big) -
 \Big[ \overline{n}_F(p)+N_B(W_{\vp+\vk})\Big] ~\delta\big(\omega+p-W_{\vp+\vk}\big)~. \label{PI1bsm}\eea

\acknowledgements This work is supported by the National Science
Foundation through the   award: PHY-0553418. D. B. thanks R.
Mohapatra and A. de Gouvea for interesting discussions.  J. Wu
acknowledges partial support through the Zaccheus Daniel Fellowship.
C. M. Ho acknowledges the support from the Croucher Foundation and
Berkeley Center for Theoretical Physics.

\end{document}